\documentclass[useAMS,usenatbib,usegraphicx,fleqn]{mn2e}
\usepackage{color}
\usepackage{aastex_hack}
\usepackage{amssymb}

\newcommand{\optprefix}[1]{(#1\mbox{-})}

\title[CRTS close supermassive black hole binaries]{A systematic search for close supermassive black hole binaries in the Catalina Real-Time Transient Survey}
\author[M. J. Graham et al.]{Matthew~J.~Graham,$^1$\thanks{E-mail:mjg@caltech.edu} 
S.~G.~Djorgovski,$^1$ Daniel Stern,$^2$ Andrew~J.~Drake,$^1$ 
\newauthor
Ashish~A.~Mahabal,$^1$  Ciro~Donalek,$^1$ Eilat~Glikman$^3$, Steve Larson$^4$, Eric Christensen$^4$\\
$^{1}$California Institute of Technology, 1200 E. California Blvd, Pasadena, CA 91125, USA \\
$^{2}$Jet Propulsion Laboratory, California Institute of Technology, 4800 Oak Grove Drive, Pasadena, CA 91109, USA \\
$^{3}$Department of Physics, Middlebury College, Middlebury, VT 05753, USA\\
$^{4}$University of Arizona, Department of Planetary Sciences, Lunar and Planetary Lab, Tucson, AZ 85721, USA
}

\begin{document}

\date{Accepted . Received ; in original form}

\pagerange{\pageref{firstpage}--\pageref{lastpage}} \pubyear{2011}

\maketitle

\label{firstpage}

\begin{abstract}
Hierarchical assembly models predict a population of supermassive black hole (SMBH) binaries. These are not resolvable by direct imaging but may be detectable via periodic variability (or nanohertz frequency gravitational waves). Following our detection of a 5.2 year periodic signal in the quasar PG 1302-102 \citep{graham15}, we present a novel analysis of the optical variability of 243,500 known spectroscopically confirmed quasars using data from the Catalina Real-time Transient Survey (CRTS) to look for close ($< 0.1$ pc) SMBH systems. Looking for a strong Keplerian periodic signal with at least 1.5 cycles over a baseline of nine years, we find a sample of 111 candidate objects. This is in conservative agreement with theoretical predictions from models of binary SMBH populations. Simulated data sets, assuming stochastic variability, also produce no equivalent candidates implying a low likelihood of spurious detections. The periodicity seen is likely attributable to either jet precession, warped accretion disks or periodic accretion associated with a close SMBH binary system. We also consider how other SMBH binary candidates in the literature appear in CRTS data and show that none of these are equivalent to the identified objects. Finally, the distribution of objects found is consistent with that expected from a gravitational wave-driven population. This implies that circumbinary gas is present at small orbital radii and is being perturbed by the black holes. None of the sources is expected to merge within at least the next century. This study opens a new unique window to study a population of close SMBH binaries that must exist according to our current understanding of galaxy and SMBH evolution.

\end{abstract}


\section{Introduction}

Supermassive black hole (SMBH) binary systems are an expected consequence of hierarchical models of galaxy formation \citep{haehnelt02, volonteri03} and an important sources of nanohertz frequency gravitational waves. Theoretically, they are seen to evolve through three stages (for a recent review, see \cite{colpi14} and references therein): in the first, the SMBH pair sinks towards the center of the newly merged system via dynamical friction (the SMBHs feel the collective effect of the stellar distribution). The binary continues to decay due to interactions with stars in the nuclear region on intersecting orbits with the binary. These stars carry away energy and angular momentum. The merger process may stall in this phase (at separations of 0.01 -- 1 pc) owing to the depletion of stars in the nuclear region. However, other factors such as the binary mass,  the presence of cold gas in the nuclear region, or deviations from axial symmetry in the galaxy remnant can affect the length of the potential stall. Depending on these factors as well as the stellar distribution, the binary will generally spend most of its lifetime in this stage. Finally, if the binary separation decreases sufficiently through these processes, the remaining orbital angular momentum of the pair is efficiently dissipated by the emission of gravitational radiation, leading to an inevitable merger. 
At coalescence, the merged SMBH receives a kick velocity and may recoil and oscillate about the core of the galaxy or even be ejected.

SMBH binary pairs have been detected at kiloparsec separations, e.g., NGC 6240 at a separation of 1.4 kpc by {\em Chandra} \citep{komossa03}, but the observational evidence for close (subparsec) pairs is tentative. The highest milliarcsecond angular resolution imaging with VLBA/VLBI  radio observations is only feasible for the very nearest systems -- a 1 pc separation pair at a distance of 100 Mpc ($z \sim 0.024$) subtends 2 milliarcsecs.  \cite{rodriguez06} report the discovery of two resolved compact, variable, flat-spectrum radio sources with a projected separation of 7.3 pc in the radio galaxy 0402+379 ($z = 0.055$). The best known close SMBH binary candidate, OJ 287 \citep{valtonen08}, shows a pair of outburst peaks in its optical light curve every 12.2 years for at least the last century, which has been interpreted as evidence for a secondary SMBH perturbing the accretion disk of the primary SBMH at regular intervals \citep{valtonen11}. Most other claims of photometric \optprefix{quasi}{periodicity}, e.g., \cite{fan02,rieger00,depaolis02,liu14}, however, tend to have far shorter temporal baselines and do not show the same level of regularity.  

To date the search for SMBH binary systems on \optprefix{sub}{parsec} scales has therefore focused on detection by spectroscopic discovery, almost solely within the SDSS data set. It should be noted, though, that there is no unique spectral characteristic signature of a SMBH binary, despite several theoretical efforts to identify one, e.g., \cite{shen10, montuori11}. Initial searches \citep{bogdanovich08, dotti09, boroson09} looked for double broad-line emission systems -- thought to originate in gas associated with the two SMBHs, where the velocity separation between the two emission line systems traces the projected orbital velocity of the binary -- or single line systems with a systematic velocity offset of the broad component from the narrow component (e.g., \cite{tsalmantza11,tsai13}), under the assumption that only one SMBH is active and Keplerian rotation about this produces the shift. However, it is possible to explain such phenomena via a disk emitter model with a single SMBH rather than requiring a binary system.  

More recent searches have been based on multi-epoch spectroscopy, looking for temporal changes in the velocity shifts consistent with binary motion (acceleration effects). This can either be from dedicated follow-up spectroscopy of initial epoch candidates (e.g., \cite{eracleous12, decarli13}) or more general searches using emerging collections of multi-epoch data, e.g., \cite{shen13, liu14, ju13}. With typical velocity offsets of $\sim1000$ km s$^{-1}$, searches are typically sensitive to systems with mass $M \sim 10^{8} M_\odot$ separated by $\sim0.1$ pc and with orbital periods of $\sim300$ years. They will therefore only capture a fraction of an orbital cycle and other astrophysical processes, such as an orbiting hotspot produced by a local instability in the broad line region (BLR) of a galaxy or outflows associated with the accretion disk, may still be responsible for any effect seen \citep{bogdanovich14, barth15}.

At small separations ($\sim$ 0.01 pc), the size of any BLR is significantly larger than the semimajor axis of the binary. The velocity dispersion of the BLR gas bound to a single SMBH is higher than the velocity difference between the two SMBHs in a binary. It is very unlikely, therefore, that emission lines from the two BLRs associated with the SBMHSs will be resolved since their separation is likely to be smaller than their widths. 
The optical and UV spectral features are therefore no longer directly related to the period of the binary \citep{shen10}. However, variability related to the periodicity of the accretion flows onto the binary may be directly detectable \citep{artymowicz96, hayasaki08}.

As noted earlier, there have been reports of \optprefix{quasi}{periodic} behavior detected in the long-term multiwavelength monitoring of particular blazars, e.g., PKS 2155-304, AO 0235+164 \citep{rani09, wang14}. These can be separated into short timescale high-energy (X-ray, gamma ray) phenomena, typically with periods of a few hours or days, and longer timescale radio and optical behaviors with timescales of a fraction of a year to decades. Given the nature of these objects, though, these are generally explained as related to jet-disk interactions around a single SMBH rather than a SMBH binary.

The expected period for a pair of $10^8 M_{\odot}$ black holes at small separations  ($\sim0.01$ pc) is $\sim$10 years and there are now a number of synoptic sky surveys with sufficient temporal baselines and sky coverage for a large scale search for such objects based on their photometry \citep{pojmanski02,udalski93,rau09,sesar11}. In fact, \cite{haiman09} proposed systematic searches for periodic quasars in large optical or X-ray surveys and calculated the likely detection rates for various observing strategies. The expectation is that such objects are rare -- \cite{volonteri09} estimate that there should be $\sim$10 with $z < 0.7$. 

In this work, we present a systematic search for periodic behavior in quasars covered by the Catalina Real-time Transient Survey (CRTS\footnote{http://crts.caltech.edu}; \cite{drake09,mahabal11,djorgovski12}). This is the largest open (publicly accessible) time domain survey currently operating, covering $\sim 33000$ deg$^2$ between $-75^{\circ} < {\mathrm Dec} < 70^{\circ}$ (but avoiding regions within $\sim 10^{\circ} - 15^{\circ}$ of the Galactic plane) to a depth of $V \sim 19$ to 21.5. Time series exist\footnote{http://www.catalinadata.data} for approximately 500 million objects with an average of $\sim$250 observations over a 9-year baseline.

This paper is structured as follows: in section 2, we present the selection techniques for identifying periodic quasars and in section 3, the data sets we have applied them to. We discuss our results in section 4. We consider the CRTS time series of other periodic candidates in the literature in section 5. We discuss our findings in section 6 and present our conclusions in section 7. We assume a standard WMAP 9-year cosmology ($\Omega_\Lambda = 0.728$, $\Omega_M = 0.272$, $H_0 = 70.4$ km s$^{-1}$ Mpc$^{-1}$; \cite{jarosik}) and our magnitudes are approximately  on the Vega system.

\section{Identifying periodic quasars}

The CRTS time series are typically noisy, gappy and irregularly sampled and this precludes any standard Fourier-based analysis of periodicity which assumes regular sampling. Variant techniques that seek to regularize the data, such as Lomb-Scargle, may be $\sim 30$\% ineffective for this type of data \citep{graham13}  and so are not suitable for this analysis. Instead we have developed a joint wavelet and autocorrelation function based approach to select our periodic quasar candidates.

\subsection{Wavelet analysis}

Wavelets are an increasingly popular tool in analyses of time series and are particularly attractive since they allow both localized time and frequency analysis. In particular, the power spectrum of a time series can be evaluated as a function of time and the time evolution of parameters associated with possible \optprefix{quasi}{periodic} behavior determined, i.e., period, amplitude and phase. Although conventional wavelet analysis via the discrete wavelet transform requires regularly sampled data, a number of techniques have been developed to deal with irregularly sampled data. 

The discrete wavelet transform (DWT) of a time series, $X_t$, is a convolution with the complex conjugate of the particular wavelet function, $f$, being used:

\[ W (\omega, \tau, X_t) = \sqrt{\omega} \sum_{\alpha = 1}^{N} X_{t_\alpha} f^{\ast}(\omega(t_\alpha - \tau))  \]

\noindent
where $\omega$ is the test frequency (scale factor) and $\tau$ is an offset relative to the start of the time series (time shift).  A popular choice for the convolution function (wavelet) is the (abbreviated) Morlet wavelet, a plane wave modulated by a Gaussian decay profile (determined by the decay constant, $c$):

\[ f(z) = \exp(i \omega(t - \tau) - c\omega^{2}(t - \tau)^2) \]

\noindent
With unevenly sampled data, however, the DWT is prone to spurious high-frequency spikes and fake structures in the time-frequency plane. \cite{foster96} proposed that the DWT can be reconsidered as a projection of the time series onto a trial function $\phi(t) = \exp(i\omega(t - \tau))$ with statistical weighting $w_a = \exp(-c\omega^2(t - \tau)^2)$, which compensates for irregularities in the data coverage. Furthermore, a bias at lower frequencies due to effectively sampling more data points through a wider window can be corrected by using $Z$-statistics to give the weighted wavelet $Z$-transform (WWZ).
 
The peak of the WWZ (for constant $\tau$) can be used to determine the period of any signal in a time series within a given time window and the tracks of peaks in the time-frequency plane will show any changes of period or amplitude with time. This also allows a check on whether any periodicity is present for the whole observed time or is transient. The resolution of WWZ is dependent on the decay constant $c$ used to determine the width of the wavelet window. If the value is too high then the WWZ power decays too rapidly and small amplitude periodic signals will tend to be missed whilst if it is too low then the WWZ only has broad frequency resolution. 
A general recommendation is to use a value such that the exponential term decreases significantly in a single cycle $2\pi / \omega$. Finally, one limitation of the WWZ is that it cannot be used for data with gaps larger than the period in question.

\begin{figure}
\caption{The characteristic restframe timescale determined from the Slepian wavelet variance as a function of absolute magnitude for the 243,500 quasars in the data set. The black points indicate the median value in each magnitude bin.}
\label{slepmag}
\includegraphics[width=3.5in]{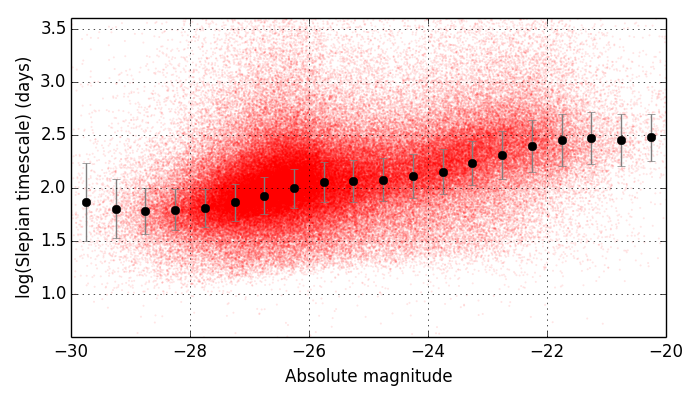}
\end{figure}

An alternate approach finds the wavelets as solutions to an eigenvalue problem where they maximize the energy in a frequency band $|f - f_c| \le f_w$ about a central frequency $\pm f_c$. These Slepian wavelets are related to discrete prolate spheroidal functions and have the advantage that they can be applied to irregular and gappy time series \citep{mondal11}. This type of wavelet analysis makes use of the variance of the wavelet coefficients to identify those scales making the largest contribution to the total wavelet variance of a time series.
\cite{graham14} used this to identify a characteristic $\sim$50 day restframe timescale in quasars that marks an apparent deviation from the expected behavior for a first-order continuous time autoregressive process (CAR(1), also known as a damped random walk or Ornstein-Uhlenbeck process). This restframe timescale also strongly (anti)correlates with absolute magnitude (see Fig.~\ref{slepmag}).

Periodic behavior is not expected as a result of a CAR(1) process (although see the discussion in 
Sec.~\ref{carma} about higher order processes). Periodic objects should therefore show a strong deviation in their wavelet variance from that expected for a CAR(1) model with a shorter characteristic timescale than regular quasars. A periodic candidate should therefore show as a statistically significant outlier from the timescale - absolute magnitude relationship, appearing below the median curve in Fig.~\ref{slepmag}.

\subsection{Autocorrelation function (ACF)}

The autocorrelation function (ACF) is a measure of how closely a quantity observed at a given time is related to the same quantity at another time (the time difference between the two is the {\em lag time}). In the case of a \optprefix{quasi}{periodic} signal, peaks in the ACF will indicate lag times at which the signal is perfectly (highly) correlated with itself and these can be interpreted as periods at which the signal \optprefix{quasi}{repeats}. The peak at zero lag $(\tau = 0)$ is the strongest peak in the ACF of a \optprefix{quasi}{periodic} signal. Successive peaks at increasing multiples of the period are increasingly weaker, and as the lag value approaches the coherence time -- the time required for a signal to change its frequency and phase information, the peaks in the autocorrelation function approach the ``noise floor'' generated by the non-periodic components of the signal and the noise.

\cite{mcquillan} discuss the robustness of the ACF for period detection in time series data. Since the ACF measures only the degree of self-similarity of the light curve at a given lag time, the period remains detectable even when the amplitude and phase of the photometric modulation evolve significantly during the timespan of the observations. The ACF is also not susceptible to residual instrumental systematics since correlated noise, long-term trends and discontinuities manifest as monotonic trends in the ACF, on top of which local maxima may still be identified. In fact, the ACF is preferable as a period-finding method to the Lomb-Scargle periodogram in terms of clarity and robustness.

For irregularly sampled data, the standard ACF estimator of \cite{edelson} can be defined as:
\[ ACF(k \Delta\tau) = \frac{\sum_{i=1}^{n}\sum_{j>i}x_i x_j b_k(t_j - t_i)}{\sum_{i=1}^{n}\sum_{j=1}^{n} b_k(t_j - t_i)} \]

\noindent
where the observations have been normalized to zero mean and unit variance before the analysis. The kernel $b_k(t_j - t_i)$ selects the observations whose time lag is not further than half the bin width from $k \Delta \tau$:

\begin{eqnarray*}
 b_k(t_j - t_i) & = &  1 \,\, \mathrm{for}  \mid (t_j - t_i) / \Delta \tau - k \mid \,\, < 1/2 \\ 
& & 0  \,\, \mathrm{otherwise} 
\end{eqnarray*}

\begin{figure*}
\caption{The autocorrelation function for the given quasar light curve (left) evaluated using the $z$-transform method (middle) and Scargle's method (right). The dashed line shows the best-fit polynomial for the $z$-transform fit constrained to go through ACF(0) = 1.}
\label{acfexample}
\includegraphics[width=7.0in]{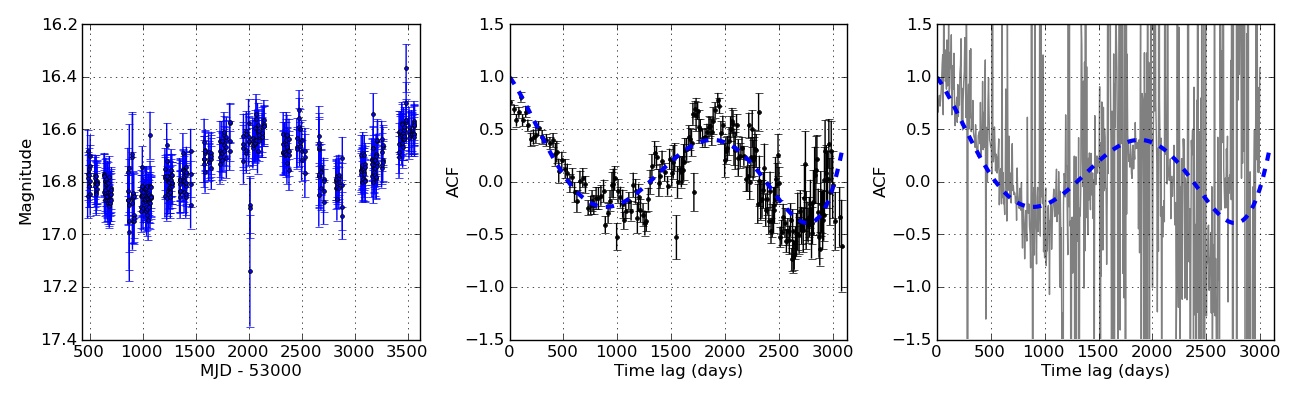}
\end{figure*}

\noindent
However, this estimator has a number of disadvantages including a high variance and not always producing positive semidefinite covariance matrix estimates (necessary to ensure the non-negativity of its Fourier transform; \cite{rehfeld}). Whilst a Gaussian kernel addresses some of these,  \cite{alexander13} recommends  Fisher's $z$-transform and equal population binning (at least 11 points in each $\Delta \tau$ bin) for a more effective estimator. Fig.~\ref{acfexample} shows the ACF for a sample quasar light curve calculated using both the z-transform method and Scargle's alternative method \citep{scargle89}.

\cite{mcquillan} identify the rotation periods of {\it Kepler} field M-dwarfs from the largest peaks (relative to neighbouring local minima) in their ACFs smoothed with a fixed width Gaussian kernel. Our data, however, have variable sampling in terms of cadence, temporal baseline and number of observations, which makes defining a single smoothing width value difficult. We have found that a better procedure to determine the period of any signal in the data is to fit the ACF of a light curve with a Gaussian process model (also known as kriging). Assuming the process has a squared-exponential covariance, this provides the best linear unbiased prediction of the underlying function from noisy data. We define the period as the time-lag value associated with the largest peak in the ACF between the second and third zero-crossings of the fitted model. This shows very good agreement with values determined from ACFs smoothed with a range of different width Gaussian kernels. We have verified the period determined by this method with that obtained with the conditional entropy algorithm of \cite{graham13}. We would expect broad agreement between the period determined from the ACF and that identified by the wavelet technique.

Periodically driven stochastic systems are expected to have an ACF with the form of an exponentially decaying cosine function \citep{jung93}. Alternatively, if the light curve is the result of a continuous autoregressive moving average CARMA process (CARMA) (see Sec.~\ref{carma}) then its ACF is a sum of exponentially damped sinusoids and exponential decays, depending on whether the roots of the characteristic polynomial of the CARMA process are real or imaginary \citep{kelly14}. To test either case, we have also determined the best fitting exponentially damped sinusoid to the ACF of the form: $ACF(\tau) = A \cos (\omega \tau) \exp(-\lambda \tau)$, where the amplitude $A$, period $p = 2 \pi / \omega$, and decay $\lambda$ are determined using maximum likelihood estimation.

\subsection{Shape and coverage}

In the simplest case, periodicity associated with a Keplerian orbit would show as a pure sinusoidal signal in the light curve. Although the projective geometry of a system -- orbital inclination, eccentricity and argument of periastron -- will distort this, the variation will remain a smooth periodic function, essentially a multiharmonic sine function. Regular flaring activity and similar phenomena not necessarily associated with a close SMBH binary pair would show as \optprefix{quasi}{periodic} discontinuities. We can therefore use the scatter around an expected sinusoidal waveform to identify those objects most closely exhibiting the desired behavior.

\begin{figure}
\caption{The median absolute deviation - magnitude relationship for the 243,500 quasars in the data set. The black points indicate the median value in each magnitude bin. The contours indicate the 1$\sigma$, 2$\sigma$ and 3$\sigma$ levels.}
\label{mad}
\includegraphics[width=3.5in]{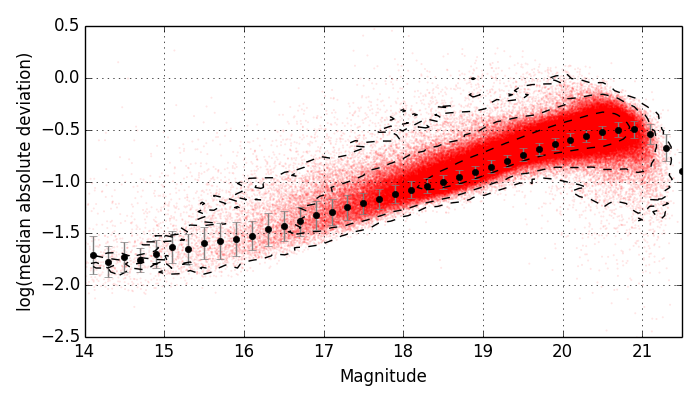}
\end{figure}

The best-fit truncated Fourier series (up to 6 terms) is determined for a given light curve phased using the period calculated by the ACF method (see above):

\[ \phi(t) = m_0 + \sum_{n=1}^{6} a_n sin (n \omega t + \phi_n) \]

\noindent
where $\phi(t)$ is the phased light curve, $m_0$ is the median magnitude and $\omega = 2 \pi / period_{ACF}$. An F-test is used to determine the exact number of harmonics to include. The scatter about the best fit will be a combination of the intrinsic variability of the quasar and the quality of the fit. We therefore scale the rms scatter of the phased data about the fit  by the median absolute deviation (MAD) of the light curve and only consider objects with scatter below some cutoff value as binary candidates. We note, however, that the MAD is a function of magnitude: fainter objects show large MAD values as there is a larger noise contribution for low S/N (faint) sources (see Fig.~\ref{mad}). We therefore employ a MAD value normalized by the median value for a given magnitude to ensure that objects with equivalent variability strength (irrespective of magnitude) can be compared.

\begin{figure}
\caption{The range of SMBH masses and binary separations to which a synoptic data set is sensitive. The solid upper line for each separation indicates a $z=5$ track and the solid lower line a $z=0.05$ track whilst the two internal dotted lines show $z=1.0$ (lower) and $z = 2.0$ (upper) tracks respectively. The hatched region indicates the range over which CRTS has temporal coverage of 1.5 cycles or more of a periodic signal. The points indicate the locations of the CRTS candidates and the best known SMBH binary candidate, OJ 287 (solid black star).}
\label{sensitive}
\includegraphics[width=3.5in]{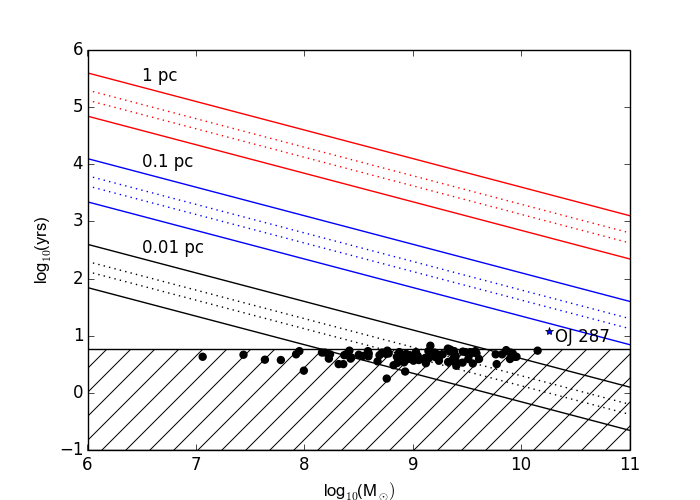}
\end{figure}

The temporal coverage of the data set used will determine the range of SMBH masses and binary separations to which it is sensitive (see Fig.~\ref{sensitive}). A decade-long baseline will only sample a fraction of an orbital period for either a low mass SMBH pair (where the primary BH mass $< 10^7 M_\odot$) 
or a more massive pair at separations $> 1$pc. Since we want objects with well-sampled periodic behavior, we limit our search to objects with coverage of at least 1.5 cycles, assuming their ACF determined period. We also only consider objects with more than 50 observations in their light curves to minimize any systematic effects in the statistics from poor sampling.

\section{Data sets}
There are few data sets with sufficient sky and/or temporal coverage and sampling to support an extensive search for quasars exhibiting periodic behavior. Most large studies of long-term quasar variability, e.g.,
SDSS with POSS \citep{mac12} or Pan-STARRS1 \citep{morganson14}, consist of relatively few epochs of data spread over a roughly decadal baseline,  which is sufficient to model ensemble behavior but not to identify specific patterns in individual objects. CRTS represents the best data currently available with which to systematically define sets of quasars with particular temporal characteristics.

\subsection{Catalina Real-time Transient Survey (CRTS)}
\label{crts}

CRTS leverages the Catalina Sky Survey data streams from three telescopes --  the 0.7 m Catalina Sky Survey Schmidt and 1.5 m Mount Lemmon Survey telescopes in Arizona and  the 0.5 m Siding Springs Survey Schmidt in Australia -- used in a search for Near-Earth Objects, operated by Lunar and Planetary Laboratory at University of Arizona. CRTS covers up to $\sim$2500 deg$^2$ per night, with 4 exposures per visit, separated by 10 min, over 21 nights per lunation. All data are automatically processed in real-time, and optical transients are immediately distributed using a variety of electronic mechanisms\footnote{http://www.skyalert.org}. The data are broadly calibrated to Johnson $V$ (see \cite{drake13} for details) and the full CRTS data set\footnote{http://nesssi.cacr.caltech.edu/DataRelease} contains time series for approximately 500 million sources. 

\begin{figure}
\caption{The $V$-band magnitude distribution for the CRTS known quasar sample (red). The tail of sources fainter than $V \sim 20.3$ are from the Mount Lemmon 1.5m telescope in the Catalina surveys. The stepped distribution (black) shows the rejected objects (less than 50 data points) from the sample which are primarily fainter sources.}
\label{magnitude}
\includegraphics[width=3.5in]{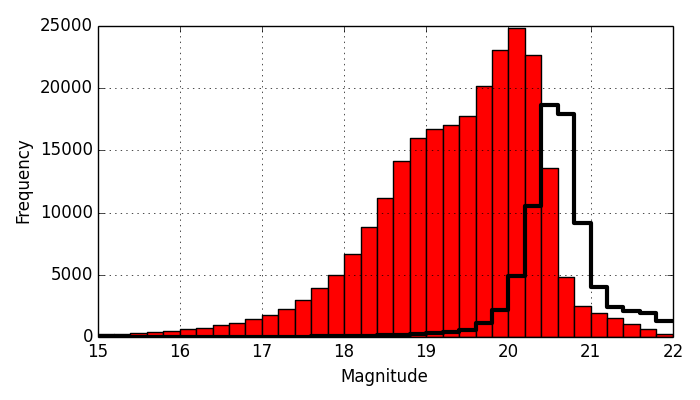}
\end{figure}

\begin{figure}
\caption{The redshift distribution for the CRTS known quasar sample (red), omitting the tail of the distribution which contains 164 quasars with $z > 5.$ The stepped distribution (black) again shows the rejected objects.}
\label{redshift}
\includegraphics[width=3.5in]{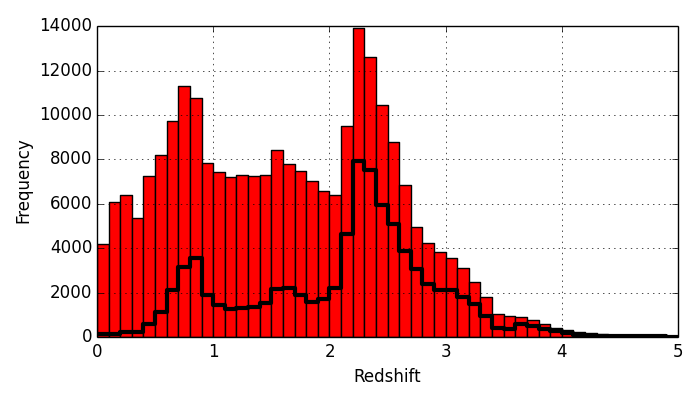}
\end{figure}

The Million Quasars (MQ) catalogue\footnote{http://quasars.org/milliquas.htm} v3.7 contains all spectroscopically confirmed type 1 QSOs (309,525), AGN (21,728) and BL Lacs (1,573) in the literature up to 2013 November 26 and formed the basis for the results of \cite{graham15}.\footnote{In this analysis we make no distinction based on object class and consider all these sources together.} We have extended this with 297,301 spectroscopically identified quasars in the SDSS Data Release 12 \citep{paris15}. We crossmatched this combined quasar list against the CRTS data set with a 3\arcsec \, matching radius and find that 334,446 confirmed quasars are covered by the full CRTS.  Of these, 83,782 do not pass our sampling criterion (i.e., they have less than 50 observations in their light curves), leaving a data set of 250,664 quasars. Note that the majority of rejected objects are faint ($V > 20$). The magnitude and redshift distributions for our sample are shown in Figs.~\ref{magnitude} and \ref{redshift}.

\subsection{Preprocessing}
It is common to preprocess data to remove spurious outlier points which may be caused by technical or photometric error. The danger, of course, is removing real signal, although a robust method should be unaffected by the presence of noisy data. We have created a cleaned version of each data set following the procedure of \cite{palanque11} (PD11): a 3-point median filter was first applied to each time series, followed by a clipping of all points that still deviated significantly from a quintic polynomial fit to the data. To ensure that not too many points are removed, the clipping threshold was initially set to 0.25 mag and then iteratively increased (if necessary) until no more than 10\% of the points were rejected. Note that PD11 used a 5$\sigma$ threshold but we found that this removed too few points and so used the limit set by \cite{schmidt10} instead.

There is always a chance that the light curve for a particular object may be contaminated by stray light (e.g., diffraction spikes, reflections) from nearby bright sources (up to $\sim$2\arcmin \, away for CRTS), low surface brightness galaxies or genuine blends, which can give a false seeing-dependent result in the analysis. If an object has a nearby counterpart (within $\sim$5\arcsec) which is unresolved by CRTS ($\sim$2 pixels), the light curve will be the combined signal from both, dominated by the brighter of the two. We have defined a distance and magnitude-based criterion to exclude quasars near bright stars by crossmatching the quasar list against the Tycho catalog of bright stars $(V < 14)$ and reject 5,319 sources. We reject all objects within a $350''$ radius of any star brighter than $V = 6.67$ and within a radius of $200'' / (V - 6)$ for $V > 6.67$. To identify blends, we crossmatched our list against SDSS and flagged those objects where there is a significant difference (taking into account the variability of the source) between the magnitude of the CRTS source and the SDSS $r$ of its match, typically indicating a faint close SDSS companion contributing to the CRTS photometry: $|V - r| > 5 \sigma_V$.  We also flagged those CRTS quasars outside the SDSS footprint (or without a SDSS companion) where the distribution of the spatial positions of individual photometry points in the light curve show significant scatter, e.g., a bimodal distribution indicating two sources: $\sigma_{pos} > 5 \sigma_pos(V)$ where $\sigma_pos(V)$ is the expected positional scatter for all quasar of magnitude $V$. Together these reject another 4.560 sources. The final size of the data set is 243.486 quasars.

\subsection{Mock data}
To first approximation, quasar variability is well-described by a Gaussian first-order continuous autoregressive model (CAR(1)), also known as a damped random walk or Ornstein-Uhlenbeck process (see \cite{graham14, kelly14} for details of where this description breaks down). Formally, the temporal behavior of the quasar flux $X(t)$ is given by:

\[ dX(t) = -\frac{1}{\tau}X(t)dt + \sigma \sqrt{dt}\epsilon(t) + b dt, \tau, \sigma, t > 0 \]

\noindent where $\tau$ is the relaxation time of the process, $\sigma$ is the variability of the time series on timescales short compared to $\tau$, $b\tau$ is the mean magnitude, and $\epsilon(t)$ is a white noise process with zero mean and variance equal to 1. For each quasar in our data set, we have generated a simulated light curve assuming that it follows a CAR(1) model. 
Using the actual observation times, $t_i$, we replace the observed magnitudes with those that would be expected under a CAR(1) model. The magnitude $X(t)$ at a given timestep $\Delta t$ from a previous value $X(t - \Delta t)$ is drawn from a Gaussian distribution with mean and variance given by, e.g., \citep{kelly09}:

\[ E(X(t) | X(t - \Delta t)) = e^{-\Delta t / \tau} X(t - \Delta t) + b\tau(1 - e^{-\Delta t / \tau}) \]

\[\mathrm{Var}(X(t) | X(t - \Delta t)) = \frac{\tau \sigma^2}{2} [1 - e^{-2\Delta t / \tau}] \]

\noindent 
We add a Gaussian deviate normalized by the photometric error associated with the magnitude to be replaced at each time $t$ to incorporate measurement uncertainties into the mock light curves. For each light curve, we set $b\tau$ to its median value and use the rest frame CAR(1) fitting functions determined by \cite{mac10}: 

\[ \log f = A + B \log \left(\frac{\lambda_{RF}}{4000 {\mathrm \AA}} \right) + C (M_i + 23) + D \log \left(\frac{M_{BH}}{10^9 M_{\odot}}\right) \]

\noindent
where $(A,B,C,D) = (-0.51, -0.479, 0.131, 0.18)$ for $f = SF_\infty$ and $(A,B,C,D) = (2.4, 0.17, 0.03, 0.21)$ for $f = \tau$. Note that a value of $C = 0.113$  is used for $f = SF_\infty$ for non K-corrected magnitudes. $M_i$ is the absolute magnitude of the quasar and $\lambda_{RF}$ is the restframe wavelength of the filter -- here taken to be SDSS $r$, $\lambda_{RF} = 6250 {\mathrm \AA} / (1 + z)$. The mass of the black hole given the absolute magnitude is drawn from a Gaussian distribution:

\[
p(\log M_{BH} | M_i) = \frac{1}{\sqrt{2\pi}\sigma} \exp \left[ - \frac{(\log M_{BH} - \mu)^2}{2\sigma^2}\right]
\] 

\noindent
where $\mu = 2.0 - 0.27 M_i$ and $\sigma = 0.58 + 0.011 M_i$. This is based on \cite{shen08} results. Differences in cosmologies used in estimating the best-fit parameters for CAR(1) and black hole mass should only have a 1\% effect (MacLeod thesis, 2012).

\section{Results}

We have applied our technique which identifies strong periodic behavior to both the real CRTS quasar data set and its simulated counterpart. From the real data set, we define an initial candidate sample according to the following criteria (see Table~\ref{selection} for a summary). For the wavelet component, we set the decay constant $c = 0.001$ and use the frequency range 0.0003 -- 0.05 day$^{-1}$. As we are interested in well-defined periodicity, we only consider those objects whose wavelet peak significance places them in the top quartile of the data set (in terms of significance), which translates to a WWZ peak value cutoff of $wwzpk > 50$. We also only consider objects with a characteristic timescale from their Slepian wavelet variance at least $1\sigma$ below the expected value for their absolute magnitude. This reflects that periodic objects are not expected to be well-described by a CAR(1) model.

\begin{table*}
\caption{The selection constraints used to define an initial sample of binary candidates and the respective counts for the real and simulated data sets for each constraint.}
\label{selection}
\begin{tabular}{llll}
\hline
Component & Constraint & Real & Mock \\
\hline
Wavelet peak value & $wwzpk > 50$ & 24437 & 32078 \\
Slepian wavelet deviation & $\tau_{slep} < \tau(M_{V}) - \sigma_{\tau}$ & 37828 & 27746 \\
WWZ-ACF period & $ 0.9 < p_{ACF}/p_{WWZ} < 1.1$ & 30330 & 63637 \\
ACF amplitude & $A > 0.3$ & 108625 &  143447 \\
ACF decay & $\lambda < 10^{-3}$ & 172049 & 182592 \\
Shape & $rms / MAD < 0.67$ & 11794 & 3944 \\
Temporal coverage & $\tau / p_{ACF} > 1.5$ & 245234 & 182257 \\
Number of points & $n > 50$ & 243486 & 243486 \\
\hline
Combined & - & 101 & 0 \\
Final & - & 111 &  0 \\
\hline
\end{tabular}
\end{table*}

The first ACF-based constraint is that the periods determined by the WWZ and the ACF, respectively, should agree to within 10 per cent. The shape of the ACF should also be reasonably described by an exponentially damped cosine with not too small an amplitude $(A > 0.3)$ and a decay constant such that the ACF drops by at most a factor of $1/e$ over the temporal baseline of the time series. This corresponds to $\lambda < 10^{-3}$ for the shortest time coverage in the data set and so we use this as a fiducial value.

In terms of the shape constraint, we limit the selection to those objects whose rms scatter about the best-fit truncated Fourier series to their phased light curve is less than the $1\sigma$ lower limit on the magnitude-normalized median absolute deviation, i.e., $rms / mad < 0.67$. Finally, as previously described, we also limit the allowed temporal coverage to 1.5 cycles or greater, i.e. $\tau / period_{ACF} > 1.5$, where $\tau$ is the timespan of a light curve, and only consider light curves with 50 or more observations.

Ideally, we want to determine the discriminating hyperplane in the multidimensional feature space that provides a clear separation between periodic and non-periodic objects. The feature cuts that we have defined provide a first approximation to this but likely candidates close to a selection value may have been missed due to error, poor sampling, etc. We therefore use the initial sample defined by the feature cuts as part of the training set for a support vector machine (SVM) with a radial basis function kernel. A number of similar machine learning algorithms could also be employed here but we found from simulations that the SVM was the best in terms of speed and accuracy.

The other part of the training set comprises a set of randomly selected quasars that do not meet the periodicity selection criteria. From simulations with known numbers of periodic objects and periodic/non-periodic ratios, we have determined that the most adequate training set mix is $\sim$100:1, given our expected detection rate (see Section~\ref{detection}). 10000 randomly selected ``non-periodic" quasars are thus added to the training set. The trained SVM is then used to classify each quasar in the full data set as periodic or not periodic based on their feature values. We repeat this process 50 times and identify those objects in the full data set that are selected each time as periodic as a final binary candidate. 

\begin{figure*}
\caption{The three objects in the simulated data which most closely match the selection criterion. The best-fit sinusoid for the determined period for each source is plotted in red.}
\label{mock}
\includegraphics[width=7.0in]{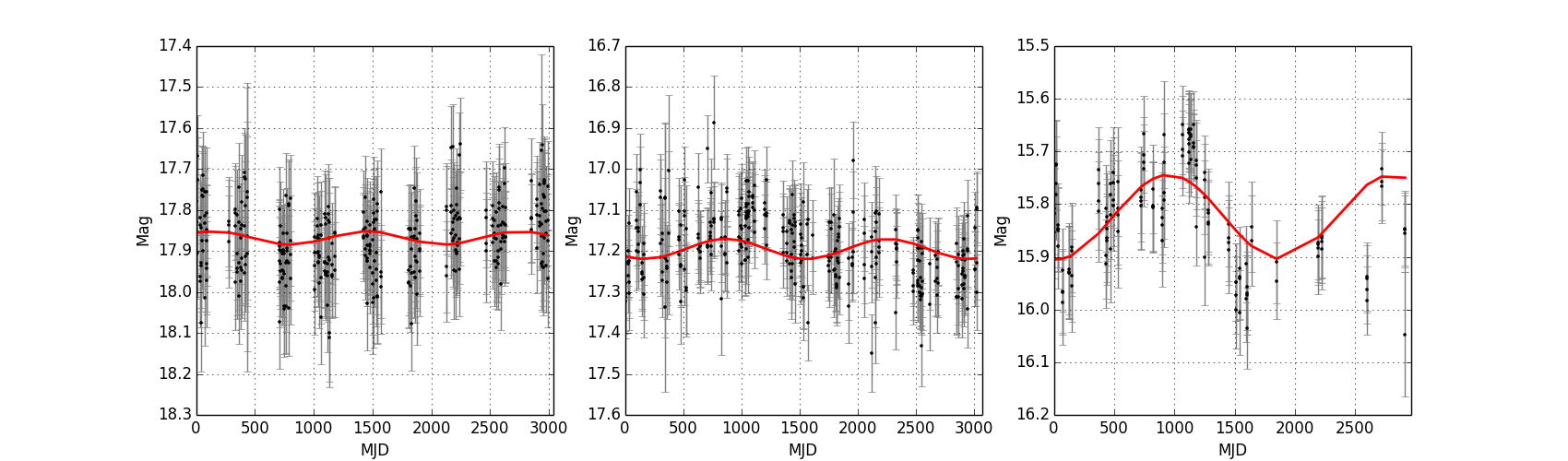}
\end{figure*}

Our final sample consists of 111 quasars (see Table~\ref{candidates} for details and Fig.~\ref{lcs} for their light curves). We have also applied the same selection process to the simulated data set of objects with CAR(1) model light curves and find that none of them satisfy our criteria. If we apply the SVM trained on the real data to the simulated data set, we also find that that no objects are selected in all iterations as periodic. There are, however, three mock sources which are selected in almost all iterations (see Fig.~\ref{mock}). Though these objects are the closest to our selection criteria, they look qualitatively different from the real sample with no clear periodicity. The range of variability they exhibit is also significantly more than in the real sample: their mock light curves have MAD values two or three times the values seen in the real data.

\begin{table*}
\caption{The 111 periodic quasar candidates meeting the selection criteria. The full version is available in electronic form online. Black hole masses are obtained from Shen et al. (2008) and updates or directly from spectra where a value is not available. Binary system parameters - separation $r$, the restframe merger time $(t_{insp})$ and the maximum induced timing residual amplitude $(\Delta t_{GW})$ - all assume $q = 0.5$.}
\label{candidates}
\begin{tabular}{llllllclll}
\hline
Id & RA & Dec & $z$ & $V_{median}$ & Period & $\log\left( \frac{M_{BH}}{M_{\odot}} \right)$ & $r$ & $t_{insp}$ & $\Delta t_{GW}$ \\
& & & & & (days) & & (pc) & (yrs) & (ns) \\
\hline
UM 211 & 00 12 10.9 & $-$01 22 07.6 & 1.998 & 17.38 & 1886 & - & - & - & - \\
UM 234 & 00 23 03.2 & +01 15 33.9 & 0.729 & 17.82 & 1818 & 9.19 & 0.013 &  $1.2 \times 10^4$ & 3.0 \\
SDSS J014350.13+141453.0 & 01 43 50.0 & +14 14 54.9 & 1.438 & 17.68 & 1538 & 9.21 & 0.009 & $2.8 \times 10^3$ & 2.4 \\
PKS 0157+011 & 02 00 03.9 & +01 25 12.6 & 1.170 & 18.04 & 1052 & - & - & - & - \\
RX J024252.3-232633 & 02 42 51.9 & $-$23 26 34.0 & 0.680 & 19.01 & 1818 & - & - & - & - \\
US 3204 & 02 49 28.9 & +01 09 25.0 & 0.954 & 18.09 & 1666 & 8.95$^{1}$ & 0.009 & $1.7 \times 10^4$ & 1.0 \\
CT 638 & 03 18 06.5 & $-$34 26 37.4 & 0.265 & 16.68 & 1515 & - & - & - & - \\
RXS J04117+1324 & 04 11 46.9 & +13 24 16.5 & 0.277 & 16.20 & 1851 & 8.16 & 0.006 & $1.4 \times 10^6$ & 0.1 \\
HS 0423+0658 & 04 26 30.2 & +07 05 30.3 & 0.170 & 15.73 & 1123 & - & - & - & - \\
2MASS J04352649-1643460 & 04 35 26.5 & $-$16 43 45.7 & 0.098 & 16.96 & 1369 & 7.78 & 0.004 & $4.1 \times 10^6$ & 0.1 \\
SDSS J072908.71+400836.6 & 07 29 08.6 & +40 08 37.0 & 0.074 & 16.08 & 1612 & 5.71 & 0.001 & $1.9 \times 10^{10} $ & 0.0 \\
SDSS J080237.60+340446.3 & 08 02 37.6 & +34 04 46.6 & 1.119 & 18.40 & 1428 & 8.96 & 0.007 & $8.8 \times 10^3 $ & 0.9 \\
SDSS J080648.65+184037.0 & 08 06 48.6 & +18 40 37.3 & 0.745 & 20.61 & 0892 & 7.99 & 0.003 & $1.7  \times 10^5 $ & 0.0 \\
SDSS J080809.56+311519.1 & 08 08 09.5 & +31 15 18.9 & 2.642 & 18.91 & 1162 & 8.36 & 0.003 & $1.2 \times 10^4 $ & 0.1 \\
SDSS J081133.43+065558.1 & 08 11 33.4 & +06 55 58.3 & 1.266 & 18.48 & 1587 & 9.39 & 0.010 & $1.8 \times 10^3 $ & 5.0 \\
SDSS J081617.73+293639.6 & 08 16 17.8 & +29 36 40.7 & 0.768 & 17.80 & 1162 & 9.77 & 0.013 & $3.6 \times 10^2 $ & 23.7 \\
FBQS J081740.1+232731 & 08 17 40.2 & +23 27 32.0 & 0.891 & 16.58 & 1190 & 9.55 & 0.011 & $7.4 \times 10^2 $ & 9.6 \\
SDSS J082121.88+250817.5 & 08 21 22.0 & +25 08 16.2 & 1.906 & 17.74 & 1886 & 9.53 & 0.011 & $8.8 \times 10^2 $ & 8.2 \\
SDSS J082716.85+490534.0 & 08 27 16.9 & +49 05 34.9 & 0.682 & 18.38 & 1612 & 8.96 & 0.009 & $2.2 \times 10^4 $ & 1.3 \\
SDSS J082827.84+400333.9 & 08 28 27.8 & +40 03 34.1 & 0.968 & 17.79 & 1886 & 8.87 & 0.008 & $3.1 \times 10^4 $ & 0.8 \\
SDSS J082926.01+180020.7 & 08 29 26.0 & +18 00 20.7 & 0.810 & 19.81 & 1449 & 8.42 & 0.005 & $1.1 \times 10^5 $ & 0.1 \\
SDSS J083349.55+232809.0 & 08 33 49.6 & +23 28 09.2 & 1.155 & 17.58 & 1086 & 9.40 & 0.008 & $7.4 \times 10^2 $ & 4.7 \\
SDSS J084146.19+503601.1 & 08 41 46.3 & +50 36 00.5 & 0.555 & 18.49 & 1694 & 7.44 & 0.003 & $1.1 \times 10^7 $ & 0.0 \\
BZQJ0842+4525 & 08 42 15.3 & +45 25 45.0 & 1.408 & 17.20 & 1886 & 9.48 & 0.012 & $1.7 \times 10^3 $ & 7.3 \\
SDSS J091554.50+352949.6 & 09 15 54.5 & +35 29 49.9 & 0.896 & 17.92 & 1369 & 9.05 & 0.008 & $7.3 \times 10^3 $ & 1.5 \\
SBS 0920+590 & 09 23 58.7 & +58 49 06.3 & 0.709 & 16.76 & 0649 & 8.76 & 0.004 & $4.1 \times 10^3 $ & 0.4 \\
SDSS J092911.35+203708.5 & 09 29 11.3 & +20 37 09.2 & 1.845 & 18.51 & 1785 & 9.92 & 0.014 & $1.8 \times 10^2 $ & 36.2 \\
HS 0926+3608 & 09 29 52.1 & +35 54 49.6 & 2.150 & 16.99 & 1562 & 9.95 & 0.012 & $8.4 \times 10^1 $ & 38.8 \\
SDSS J093819.25+361858.7 & 09 38 19.3 & +36 18 58.9 & 1.677 & 18.80 & 1265 & 9.32 & 0.007 & $8.3 \times 10^2 $ & 3.3 \\
SDSS J094450.76+151236.9 & 09 44 50.7 & +15 12 37.5 & 2.118 & 17.72 & 1428 & 9.61 & 0.009 & $2.6 \times 10^2 $ & 10.0 \\
SDSS J094715.56+631716.4 & 09 47 15.6 & +63 17 17.3 & 0.487 & 16.10 & 1724 & 9.22 & 0.012 & $1.3 \times 10^4 $ & 4.3 \\
HS 0946+4845 & 09 50 00.7 & +48 31 29.9 & 0.589 & 16.87 & 1587 & 8.59 & 0.007 & $1.0 \times 10^5 $ & 0.3 \\
KUV 09484+3557 & 09 51 23.9 & +35 42 49.2 & 0.398 & 17.77 & 1162 & 8.31 & 0.005 & $1.8 \times 10^5 $ & 0.1 \\
SDSS J102255.21+172155.7 & 10 22 55.2 & +17 21 56.0 & 1.062 & 18.48 & 1666 & 8.69 & 0.006 & $3.9 \times 10^4 $ & 0.4 \\
SDSS J102349.38+522151.2 & 10 23 49.5 & +52 21 51.8 & 0.955 & 16.96 & 1785 & 9.59 & 0.014 & $1.8 \times 10^3 $ & 12.1 \\
RXS J10304+5516 & 10 30 25.0 & +55 16 23.4 & 0.435 & 16.74 & 1515 & 8.43 & 0.006 & $2.2 \times 10^5 $ & 0.2 \\
SDSS J103111.52+491926.5 & 10 31 11.5 & +49 19 27.2 & 1.203 & 17.64 & 1612 & 9.04 & 0.008 & $7.8 \times 10^3 $ & 1.3 \\
SDSS J104430.25+051857.2 & 10 44 30.3 & +05 18 56.8 & 0.905 & 17.33 & 1333 & 9.24 & 0.009 & $3.3 \times 10^3 $ & 3.0 \\
SDSS J104758.34+284555.8 & 10 47 58.3 & +28 45 56.2 & 0.792 & 19.09 & 1851 & 8.72 & 0.008 & $6.9 \times 10^4 $ & 0.5 \\
SDSS J104941.01+085548.4 & 10 49 41.0 & +08 55 48.5 & 1.185 & 17.87 & 1428 & 9.37 & 0.009 & $1.7 \times 10^3 $ & 4.5 \\
MS 10548-0335 & 10 57 22.3 & --3 51 31.3 & 0.555 & 17.67 & 0892 & - & -  & - & - \\
CSO 67 & 11 03 27.5 & +29 48 11.2 & 0.909 & 17.52 & 2083 & 9.35 & 0.013 & $7.1 \times 10^3 $ & 5.2 \\
SDSS J110554.78+322953.7 & 11 05 54.8 & +32 29 54.1 & 0.151 & 17.45 & 1724 & 8.24 & 0.007 & $1.2 \times 10^6 $ & 0.2 \\
SDSS J113050.21+261211.4 & 11 30 50.2 & +26 12 11.8 & 1.012 & 17.69 & 2173 & 9.32 & 0.013 & $7.6 \times 10^3 $ & 4.6 \\
SDSS J113916.47+254412.6 & 11 39 16.4 & +25 44 13.0 & 1.012 & 17.61 & 2439 & 9.16 & 0.012 & $1.9 \times 10^4 $ & 2.6 \\
SDSS J114438.34+262609.4 & 11 44 38.3 & +26 26 10.1 & 0.974 & 18.39 & 1315 & 9.38 & 0.010 & $1.7 \times 10^3 $ & 4.9 \\
SDSS J114749.70+163106.7 & 11 47 49.7 & +16 31 06.8 & 0.554 & 18.78 & 1449 & 8.22 & 0.005 & $3.5 \times 10^5 $ & 0.1 \\
SDSS J114857.33+160023.1 & 11 48 57.4 & +16 00 22.7 & 1.224 & 18.14 & 1851 & 9.90 & 0.017 & $4.1 \times 10^2 $ & 37.6 \\
SDSS J115141.81+142156.6 & 11 51 41.8 & +14 21 57.0 & 1.002 & 18.11 & 1492 & 9.11 & 0.009 & $6.3 \times 10^3 $ & 1.8 \\
SDSS J115346.39+241829.4 & 11 53 46.4 & +24 18 29.8 & 0.753 & 18.41 & 1666 & 8.97 & 0.009 & $2.1 \times 10^4 $ & 1.3 \\
SDSS J121018.34+015405.9 & 12 10 18.3 & +01 54 06.2 & 0.216 & 16.90 & 1612 & 8.54 & 0.008 & $2.6 \times 10^5 $ & 0.6 \\
SDSS J121018.66+185726.0 & 12 10 18.7 & +18 57 27.0 & 1.516 & 17.42 & 1754 & 9.53 & 0.011 & $1.1 \times 10^3 $ & 8.4 \\
SDSS J121056.83+231912.5 & 12 10 56.8 & +23 19 13.0 & 1.260 & 18.47 & 1785 & 8.78 & 0.007 & $2.7 \times 10^4 $ & 0.5 \\
\hline
\end{tabular}
\end{table*}

\begin{table*}
\contcaption{}
\begin{tabular}{llllllclll}
\hline
Id & RA & Dec & $z$ & $V_{median}$ & Period & $\log\left( \frac{M_{BH}}{M_{\odot}} \right)$ & $r$ & $t_{insp}$ & $\Delta t_{GW}$ \\
& & & & & (days) & & (pc) & (yrs) & (ns) \\
\hline
SDSS J121457.39+132024.3 & 12 14 57.4 & +13 20 24.5 & 1.494 & 18.59 & 1923 & 9.46 & 0.011 & $1.8 \times 10^3 $ & 6.6 \\
SDSS J123147.27+101705.3 & 12 31 47.3 & +10 17 05.4 & 1.733 & 18.83 & 1851 & 9.20 & 0.009 & $3.5 \times 10^3 $ & 2.3 \\
SDSS J123821.84+030024.2 & 12 38 21.8 & +03 00 24.6 & 0.380 & 18.46 & 1250 & 8.92 & 0.008 & $2.2 \times 10^4 $ & 1.4 \\
SDSS J124044.49+231045.8 & 12 40 44.5 & +23 10 46.1 & 0.722 & 18.40 & 1428 & 8.94 & 0.008 & $1.6 \times 10^4 $ & 1.1 \\
SDSS J124119.04+203452.7 & 12 41 19.0 & +20 34 53.4 & 1.492 & 18.44 & 1219 & 9.40 & 0.008 & $6.8 \times 10^2 $ & 4.5 \\
SDSS J124157.90+130104.1 & 12 41 57.9 & +13 01 04.7 & 1.227 & 18.83 & 1538 & 8.95 & 0.007 & $9.4 \times 10^3 $ & 0.9 \\
PGC 3096192 & 12 50 29.0 & +06 36 11.1 & 0.133 & 16.93 & 1562 & 7.06 & 0.003 & $8.5 \times 10^7 $ & 0.00 \\
SDSS J125414.23+131348.1 & 12 54 14.2 & +13 13 48.4 & 0.655 & 18.11 & 1754 & 8.94 & 0.009 & $3.2 \times 10^4 $ & 1.2 \\
SDSS J130040.62+172758.4 & 13 00 40.6 & +17 27 58.5 & 0.863 & 19.26 & 1818 & 8.88 & 0.008 & $3.2 \times 10^4 $ & 0.8 \\
BZQJ1305-1033 & 13 05 33.0 & --10 33 19.1 & 0.286 & 14.99 & 1694 & 8.50 & 0.008 & $3.0 \times 10^5 $ & 0.4 \\
SNU J13120+0641 & 13 12 04.7 & +06 41 07.6 & 0.242 & 15.82 & 1492 & 9.14 & 0.012 & $2.0 \times 10^4 $ & 5.1 \\
SDSS J131706.19+271416.7 & 13 17 06.2 & +27 14 16.7 & 2.672 & 17.83 & 1666 & 9.92 & 0.011 & $7.7 \times 10^1 $ & 33.9 \\
SDSS J131909.08+090814.7 & 13 19 09.1 & +09 08 15.1 & 0.882 & 18.01 & 1298 & 8.67 & 0.006 & $2.8 \times 10^4 $ & 0.3 \\
SDSS J132103.41+123748.2 & 13 21 03.4 & +12 37 48.1 & 0.687 & 18.62 & 1538 & 8.91 & 0.008 & $2.4 \times 10^4 $ & 1.0 \\
SDSS J133127.31+182416.9 & 13 31 27.3 & +18 24 17.1 & 0.938 & 17.44 & 1639 & 9.39 & 0.011 & $3.0 \times 10^3 $ & 5.6 \\
SDSS J133516.17+183341.4 & 13 35 16.1 & +18 33 41.8 & 1.192 & 17.75 & 1724 & 9.76 & 0.015 & $6.0 \times 10^2 $ & 21.7 \\
SDSS J133631.45+175613.8 & 13 36 31.4 & +17 56 14.1 & 0.421 & 18.20 & 1562 & 9.03 & 0.010 & $2.4 \times 10^4 $ & 2.2 \\
SDSS J133654.44+171040.3 & 13 36 54.4 & +17 10 40.8 & 1.231 & 17.96 & 1408 & 9.24 & 0.008 & $2.5 \times 10^3 $ & 2.7 \\
SDSS J133807.69+360220.3 & 13 38 07.7 & +36 02 20.3 & 1.196 & 18.44 & 1960 & 9.14 & 0.010 & $9.1 \times 10^3 $ & 2.1 \\
SDSS J134820.42+194831.5 & 13 48 20.4 & +19 48 31.9 & 0.594 & 18.34 & 1388 & 7.63 & 0.003 & $2.8 \times 10^6 $ & 0.0 \\
SDSS J134855.27-032141.4 & 13 48 55.3 & --3 21 41.4 & 2.099 & 17.19 & 1428 & 9.89 & 0.011 & $8.7 \times 10^1 $ & 30.0 \\
SDSS J135225.80+132853.2 & 13 52 25.8 & +13 28 53.3 & 0.402 & 16.44 & 1754 & 8.75 & 0.009 & $1.0 \times 10^5 $ & 0.8 \\
SDSS J140600.26+013252.2 & 14 06 00.3 & +01 32 52.4 & 0.454 & 18.00 & 2000 & 8.41 & 0.007 & $4.7 \times 10^5 $ & 0.2 \\
SDSS J140704.43+273556.6 & 14 07 04.5 & +27 35 56.3 & 2.222 & 17.32 & 1562 & 9.94 & 0.012 & $8.4 \times 10^1 $ & 36.3 \\
HE 1408-1003 & 14 10 40.3 & --10 17 29.7 & 0.882 & 17.01 & 1923 & - & - & - & - \\
SDSS J141425.92+171811.2 & 14 14 25.9 & +17 18 11.6 & 0.409 & 18.33 & 1785 & 8.59 & 0.008 & $1.9 \times 10^5 $ & 0.4 \\
3C 298.0 & 14 19 08.2 & +06 28 35.1 & 1.437 & 16.53 & 1960 & 9.57 & 0.013 & $1.3 \times 10^3 $ & 10.5 \\
SDSS J142301.96+101500.1 & 14 23 02.0 & +10 15 00.0 & 1.052 & 17.96 & 1234 & 9.46 & 0.010 & $9. \times 10^2 $ & 6.3 \\
SDSS J143621.29+072720.8 & 14 36 21.3 & +07 27 21.1 & 0.889 & 17.07 & 1886 & 9.03 & 0.010 & $1.9 \times 10^4 $ & 1.5 \\
SDSS J143820.60+055447.9 & 14 38 20.6 & +05 54 48.1 & 0.614 & 18.71 & 1851 & 8.19 & 0.006 & $6.9 \times 10^5 $ & 0.1 \\
SDSS J144754.62+132610.0 & 14 47 54.6 & +13 26 10.4 & 0.572 & 18.39 & 1960 & 8.58 & 0.008 & $1.9 \times 10^5 $ & 0.4 \\
SDSS J144755.57+100040.0 & 14 47 55.6 & +10 00 40.4 & 0.678 & 17.65 & 0862 & 8.93 & 0.006 & $4.7 \times 10^3 $ & 0.9 \\
SDSS J150450.16+012215.5 & 15 04 50.2 & +01 22 15.8 & 0.967 & 17.75 & 1724 & 9.20 & 0.010 & $7.0 \times 10^3 $ & 2.7 \\
QNZ3:54 & 15 18 06.6 & +01 31 34.9 & 1.402 & 18.64 & 1724 & 9.27 & 0.009 & $3.1 \times 10^3 $ & 3.2 \\
SDSS J152035.23+095925.2 & 15 20 35.2 & +09 59 25.7 & 1.049 & 18.88 & 1204 & 9.12 & 0.007 & $3.3 \times 10^3 $ & 1.7 \\
SDSS J152157.02+181018.6 & 15 21 57.0 & +18 10 19.2 & 0.731 & 18.23 & 1960 & 7.95 & 0.005 & $1.6 \times 10^6 $ & 0.0 \\
SDSS J153636.22+044127.0 & 15 36 36.2 & +04 41 26.9 & 0.379 & 16.75 & 1111 & 8.82 & 0.007 & $2.4 \times 10^4 $ & 0.9 \\
SDSS J154409.61+024040.0 & 15 44 09.6 & +02 40 39.8 & 0.964 & 18.72 & 2000 & 8.76 & 0.008 & $5.5 \times 10^4 $ & 0.5 \\
SDSS J155449.11+084204.8 & 15 54 49.1 & +08 42 05.4 & 0.786 & 17.03 & 1562 & 8.85 & 0.008 & $2.6 \times 10^4 $ & 0.8 \\
SDSS J155647.78+181531.5 & 15 56 47.8 & +18 15 32.1 & 1.502 & 18.79 & 1428 & 9.51 & 0.010 & $6.7 \times 10^2 $ & 7.2 \\
SDSS J160730.33+144904.3 & 16 07 30.3 & +14 49 04.2 & 1.800 & 17.57 & 1724 & 9.82 & 0.013 & $2.5 \times 10^2 $ & 24.6 \\
SDSS J161013.67+311756.4 & 16 10 13.7 & +31 17 56.6 & 0.245 & 17.19 & 1724 & 7.92 & 0.005 & $3.2 \times 10^6 $ & 0.1 \\
SDSS J161854.64+230859.1 & 16 18 54.6 & +23 08 59.3 & 0.561 & 18.22 & 1666 & 8.37 & 0.006 & $2.9 \times 10^5 $ & 0.1 \\
163107.34+560905.3 & 16 31 07.4 & +56 09 05.1 & 0.731 & 19.13 & 1724 & 8.38 & 0.006 & $2.2 \times 10^5 $ & 0.1 \\
HS 1630+2355 & 16 33 02.7 & +23 49 28.8 & 0.821 & 15.23 & 2040 & 9.86 & 0.020 & $1.1 \times 10^3 $ & 38.7 \\
SDSS J164452.71+430752.2 & 16 44 52.7 & +43 07 52.9 & 1.715 & 17.06 & 2000 & 10.15 & 0.019 & $1.1 \times 10^2 $ & 91.6 \\
SDSS J165136.76+434741.3 & 16 51 36.8 & +43 47 41.9 & 1.604 & 18.13 & 1923 & 9.34 & 0.010 & $2.6 \times 10^3 $ & 4.1 \\
MCG 5-40-026 & 17 01 07.8 & +29 24 24.6 & 0.036 & 14.65 & 1408 & - & - & - &- \\
SDSS J170616.24+370927.0 & 17 06 16.2 & +37 09 27.0 & 1.267 & 18.19 & 1388 & 9.10 & 0.007 & $3.9 \times 10^3 $ & 1.6 \\
HS 1715+2131 & 17 17 20.1 & +21 28 15.0 & 0.590 & 16.19 & 1250 & - & - & - & - \\
FBQS J17239+3748 & 17 23 54.3 & +37 48 41.7 & 0.828 & 17.55 & 1960 & 9.38 & 0.013 & $6.0 \times 10^3 $ & 6.1 \\
SDSS J172656.96+600348.5 & 17 26 56.9 & +60 03 49.1 & 0.991 & 18.50 & 1923 & 9.15 & 0.010 & $1.1 \times 10^4 $ & 2.3 \\
4C 50.43 & 17 31 03.7 & +50 07 35.7 & 1.111 & 16.74 & 1075 & - & - & - & - \\
BZQJ2156-2012 & 21 56 33.7 & --20 12 30.2 & 1.309 & 17.01 & 1333 & - & - & - & - \\
SDSS J221016.97+122213.9 & 22 10 17.0 & +12 22 14.0 & 0.717 & 18.32 & 1333 & 9.00 & 0.008 & $1.1 \times 10^4 $ & 1.4 \\
6QZ J221925.1-305408 & 22 19 25.2 & --30 54 08.1 & 0.579 & 19.15 & 1408 & - & - & - & - \\
HS 2219+1944 & 22 22 21.1 & +19 59 48.1 & 0.211 & 16.33 & 1724 & - & - & - & - \\
SDSS J224829.47+144418.0 & 22 48 29.4 & +14 44 18.4 & 0.424 & 18.82 & 1219 & 8.86 & 0.008 & $2.4 \times 10^4 $ & 1.0 \\
\hline
\end{tabular}
\end{table*}

If we take the simulated data results as an indication of the expected number of quasars to show comparable periodic behavior by chance then the number of real objects selected is statistically significant, particularly with the higher temporal coverage constraint. It also suggests strong periodicity (on an approximately decadal timescale) is not common in quasars nor is it expected as an artifact of a CAR(1) process. We note, however, that our selection criteria will select against longer periodic as well as more general quasi-periodic behavior as may be expected from SMBH binaries at greater than 1 pc separation.

\begin{figure*}
\centering
\caption{Light curves for periodic candidates. CRTS data (black) is shown; complementary data (blue) from the LINEAR survey \citep{sesar11} is also included where available. The best-fit sinusoid for the determined period is plotted in red. The full set is available online.}
\label{lcs}
\includegraphics[width = 7.0in]{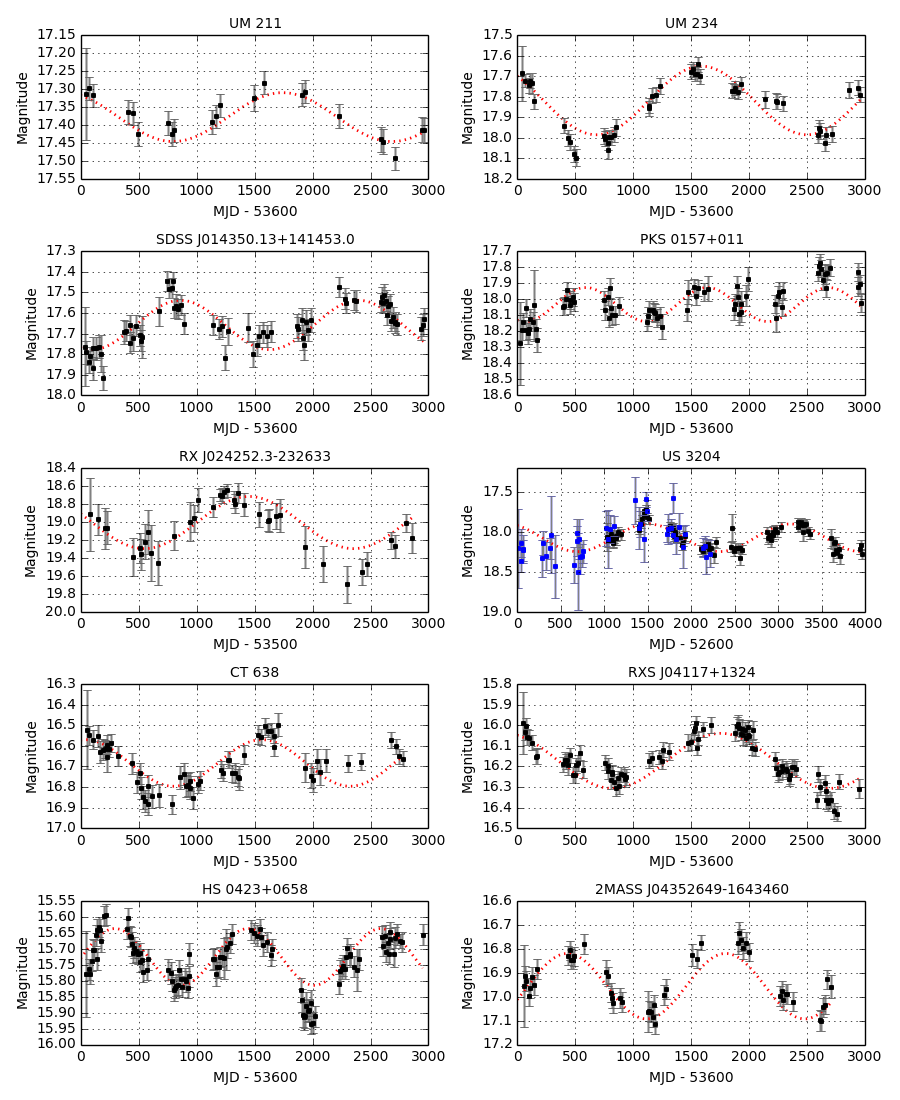} 
\end{figure*}

\begin{figure*}
\centering
\contcaption{}
\includegraphics[width = 7.0in]{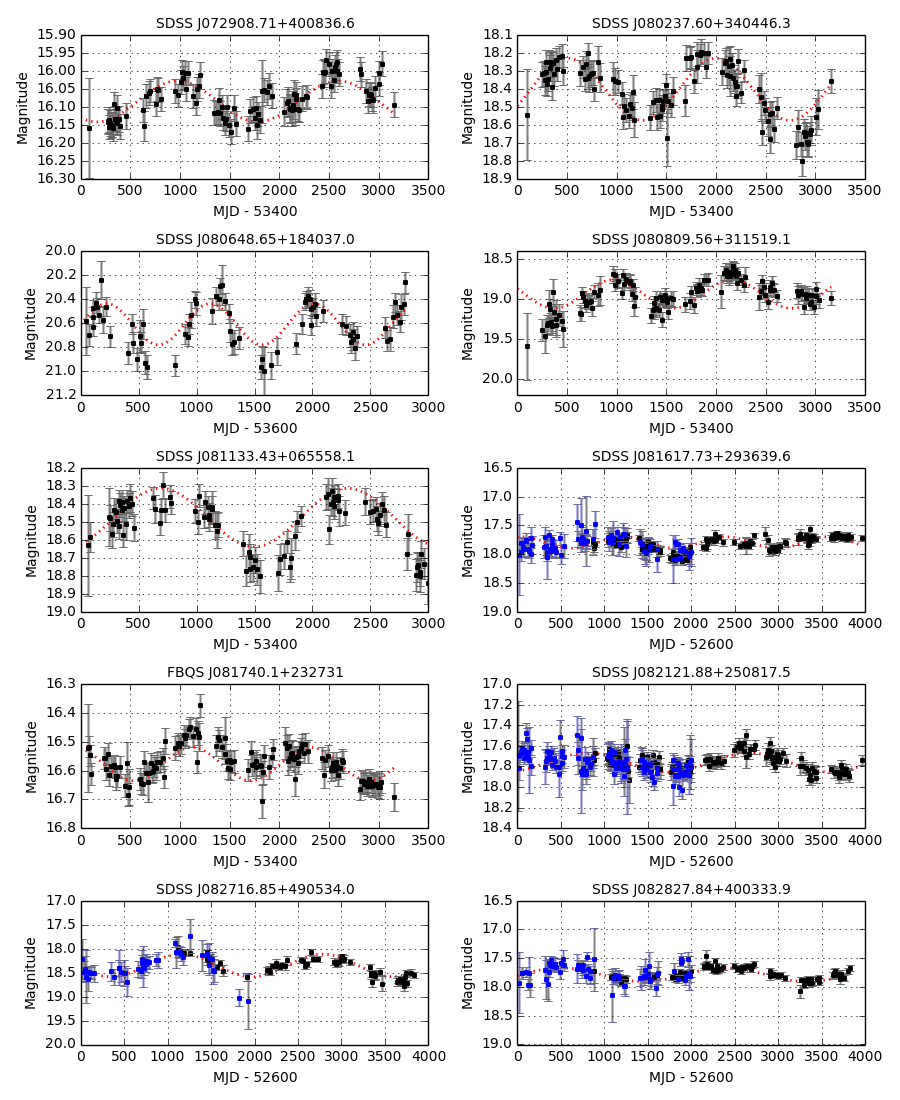} 
\end{figure*}

\begin{figure*}
\centering
\contcaption{}
\includegraphics[width = 7.0in]{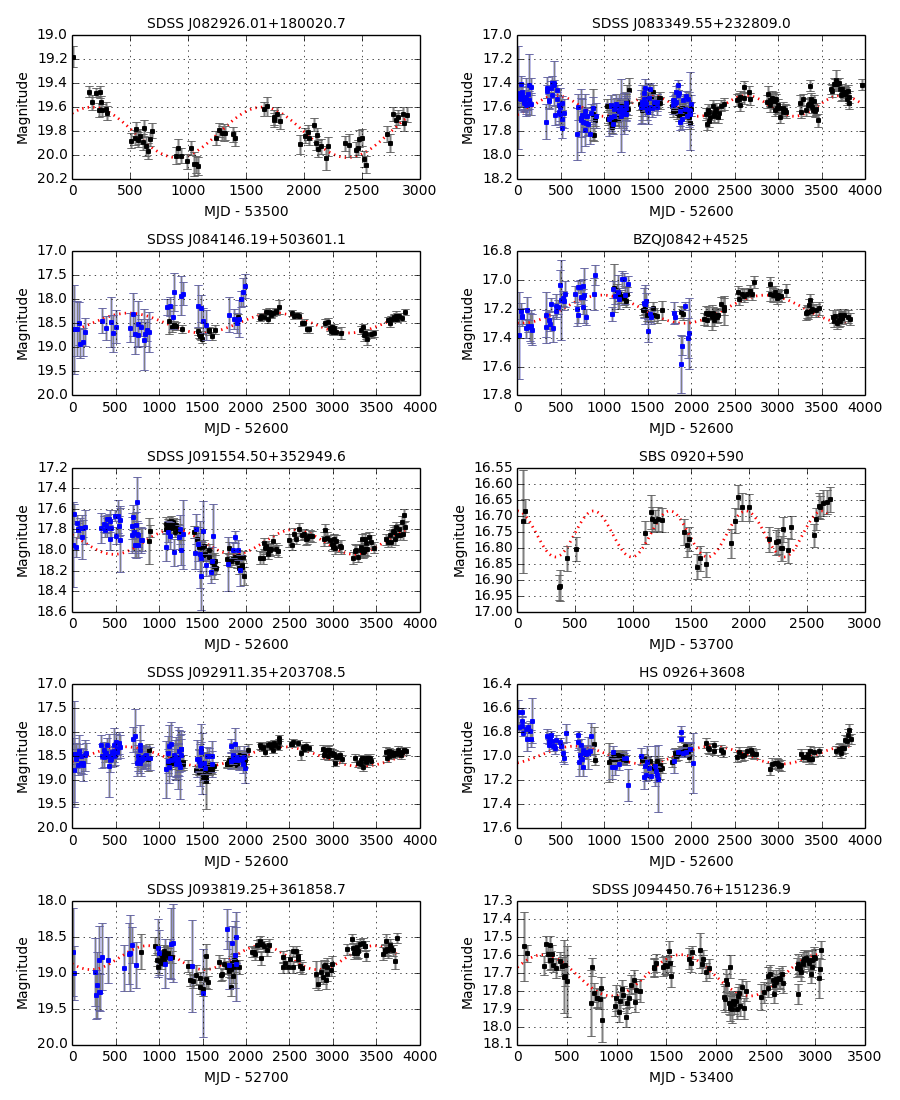} 
\end{figure*}

\begin{figure*}
\centering
\contcaption{}
\includegraphics[width = 7.0in]{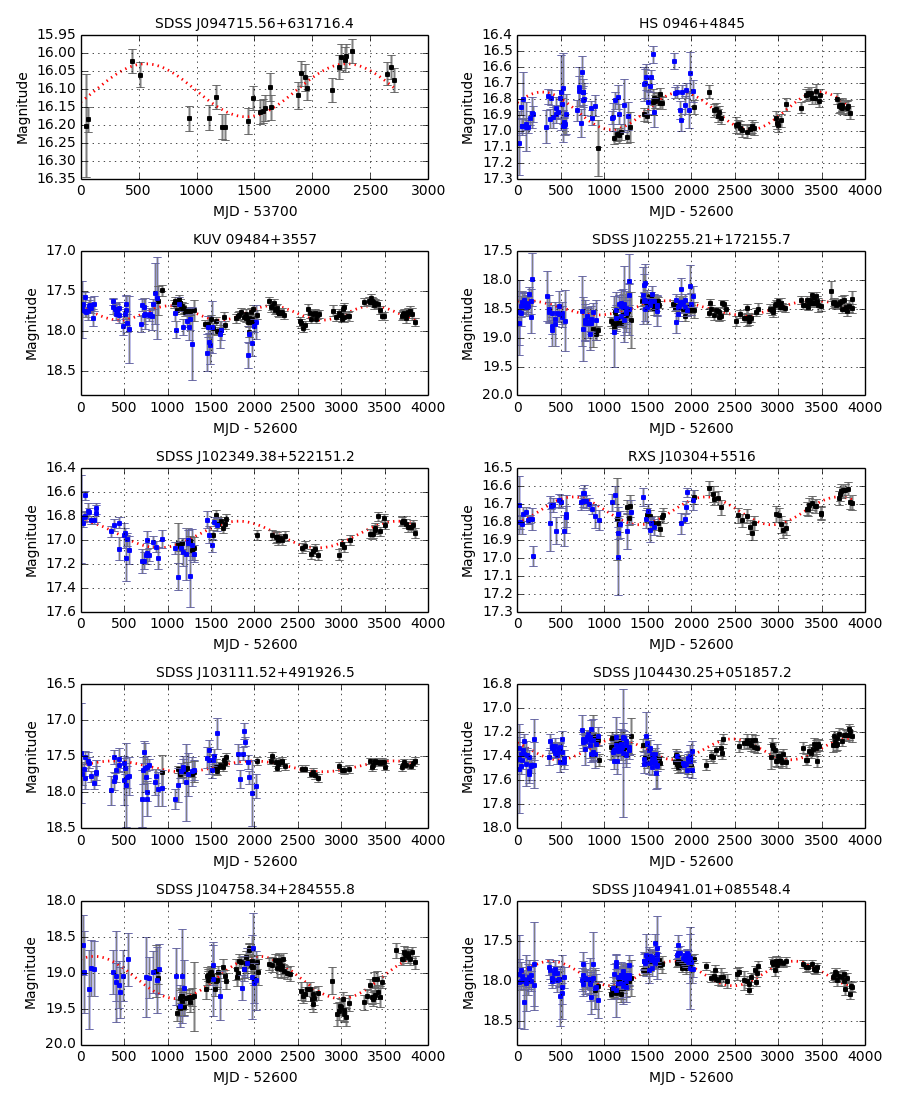} 
\end{figure*}

\begin{figure*}
\centering
\contcaption{}
\includegraphics[width = 7.0in]{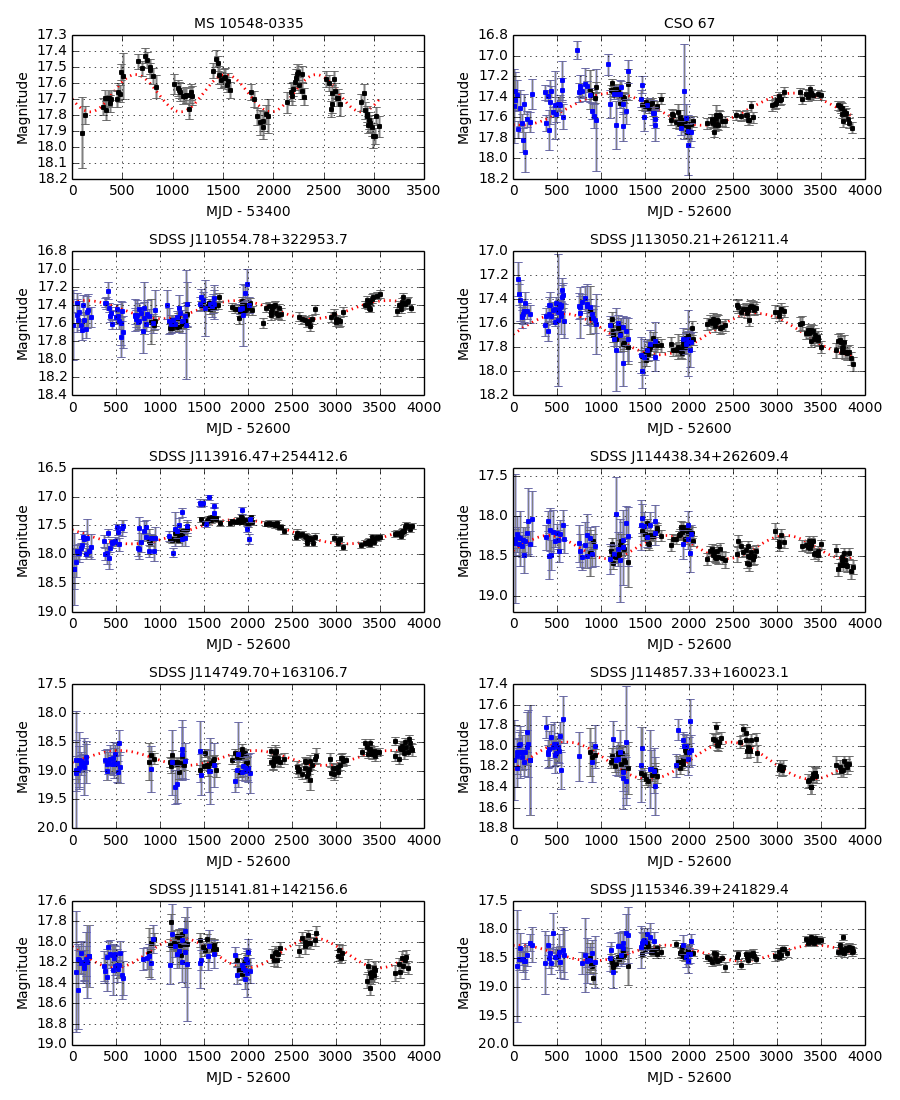} 
\end{figure*}

\begin{figure*}
\centering
\contcaption{}
\includegraphics[width = 7.0in]{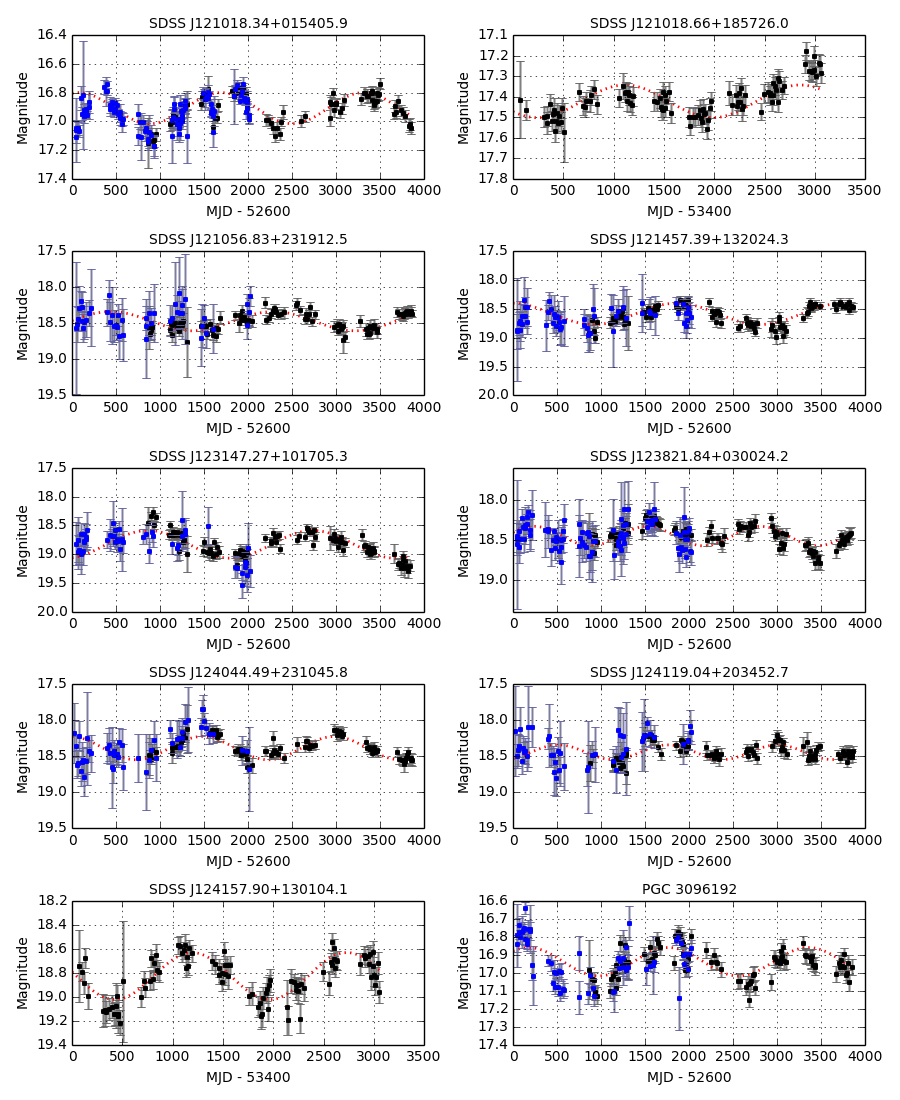} 
\end{figure*}

\begin{figure*}
\centering
\contcaption{}
\includegraphics[width = 7.0in]{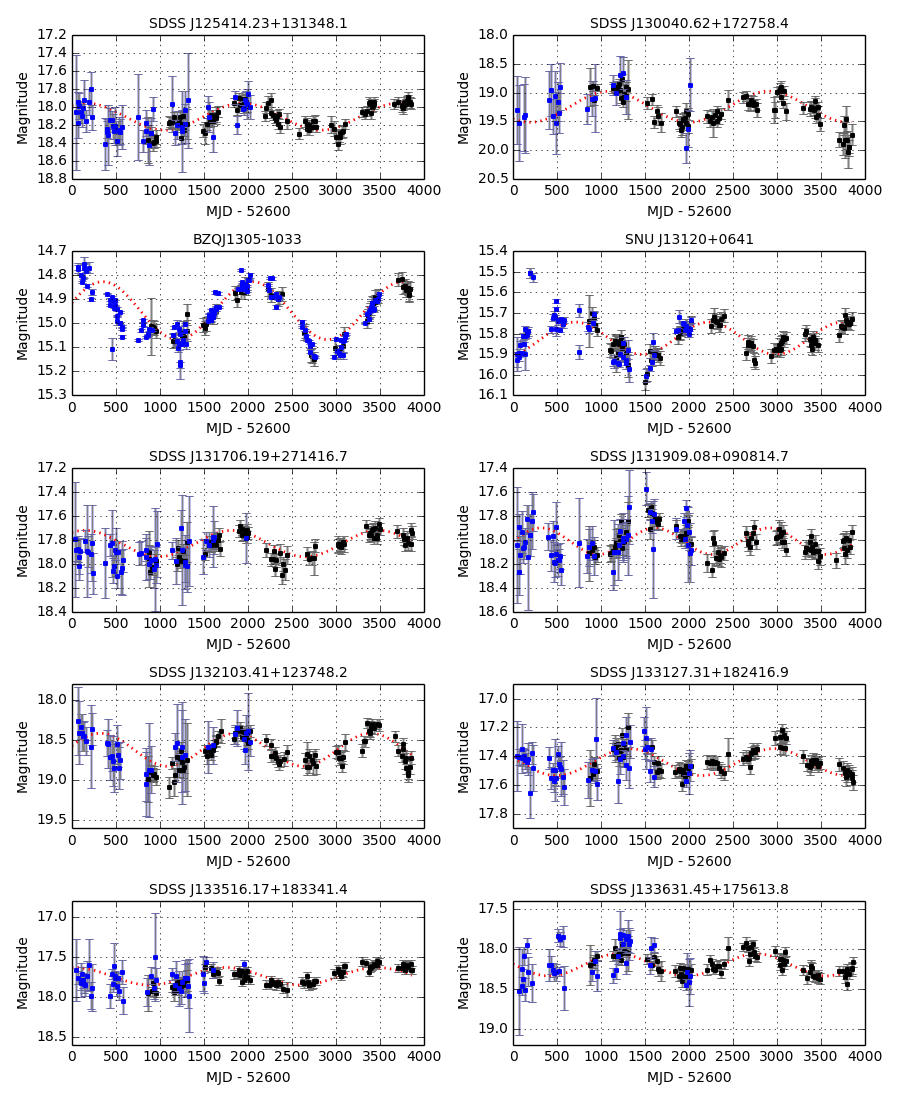} 
\end{figure*}

\begin{figure*}
\centering
\contcaption{}
\includegraphics[width = 7.0in]{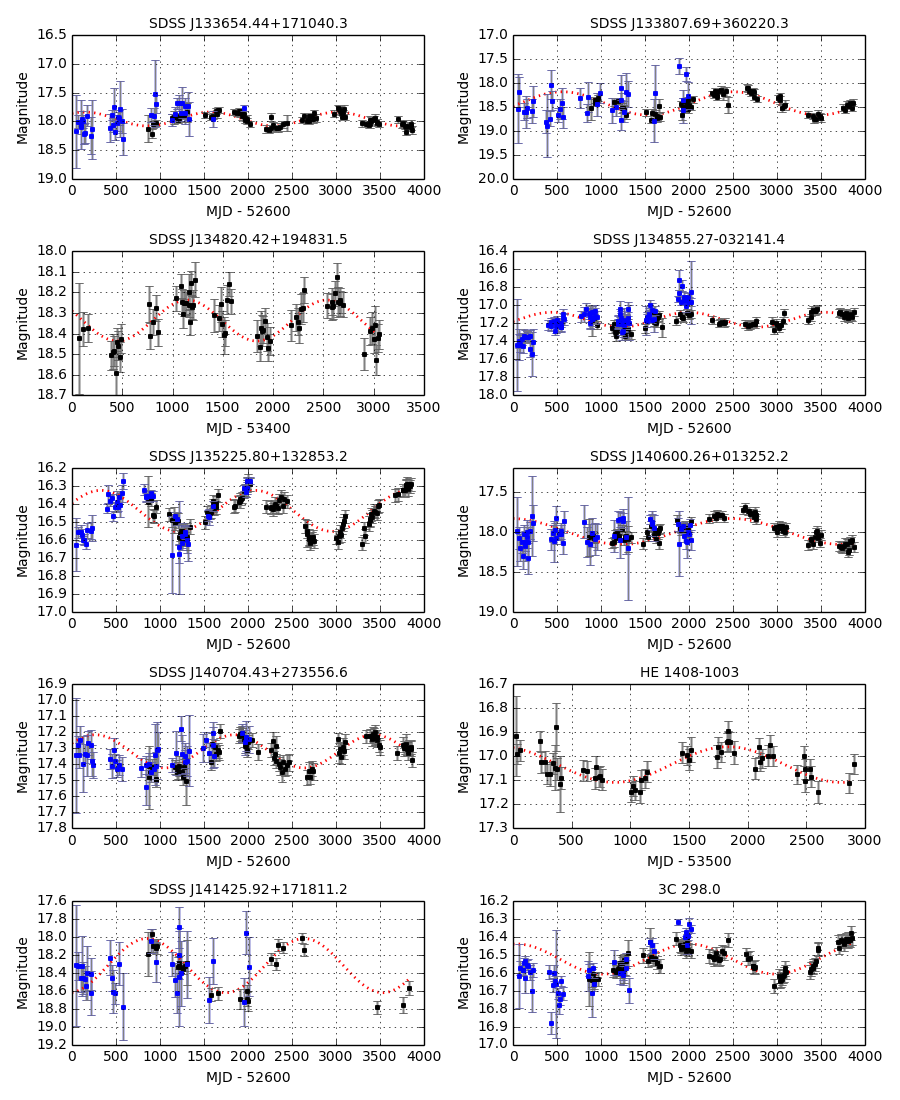} 
\end{figure*}

\begin{figure*}
\centering
\contcaption{}
\includegraphics[width = 7.0in]{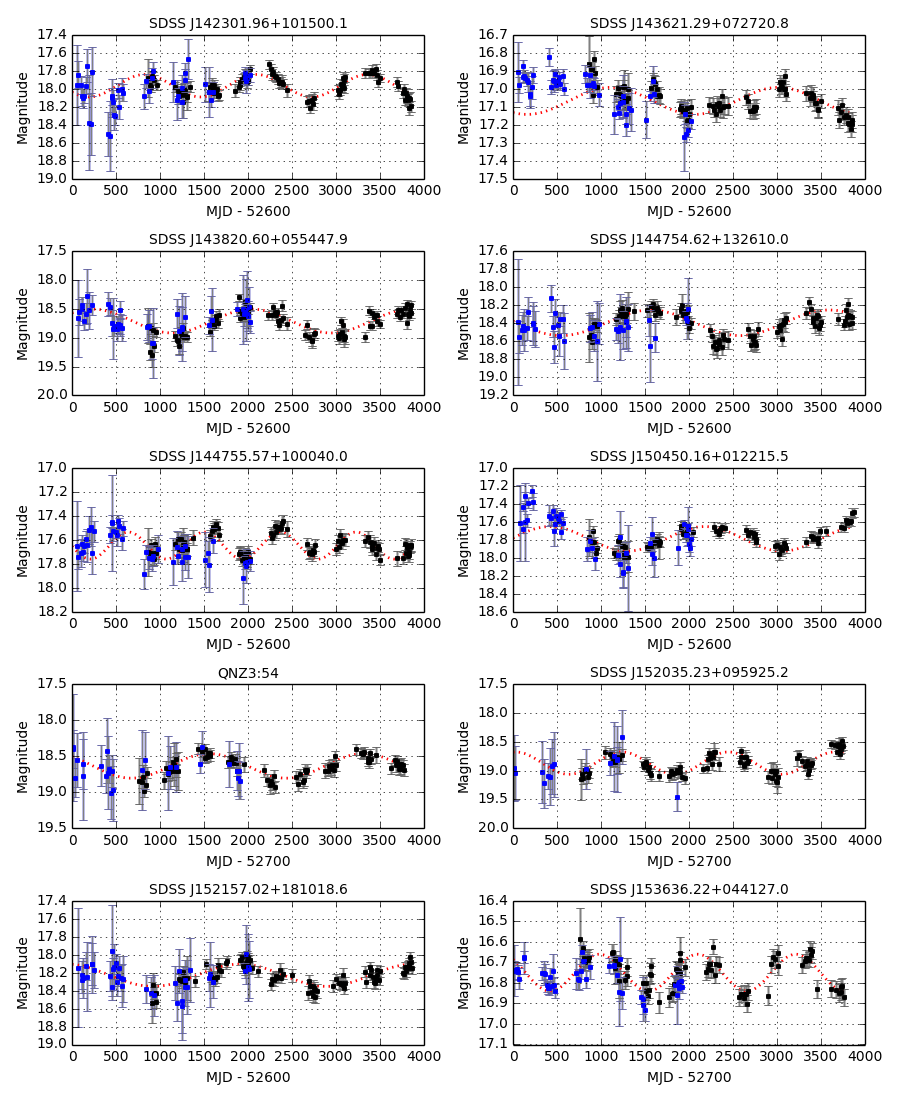} 
\end{figure*}

\begin{figure*}
\centering
\contcaption{}
\includegraphics[width = 7.0in]{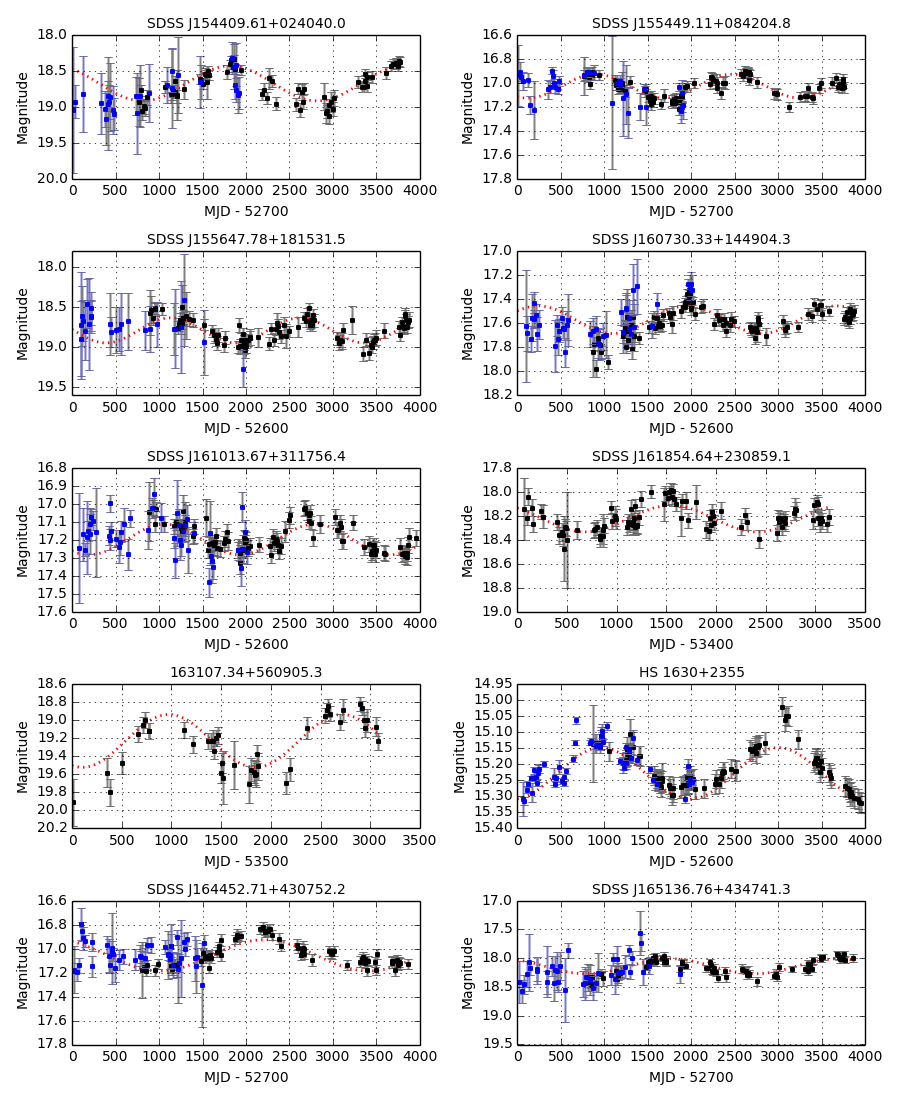} 
\end{figure*}

\begin{figure*}
\centering
\contcaption{}
\includegraphics[width = 7.0in]{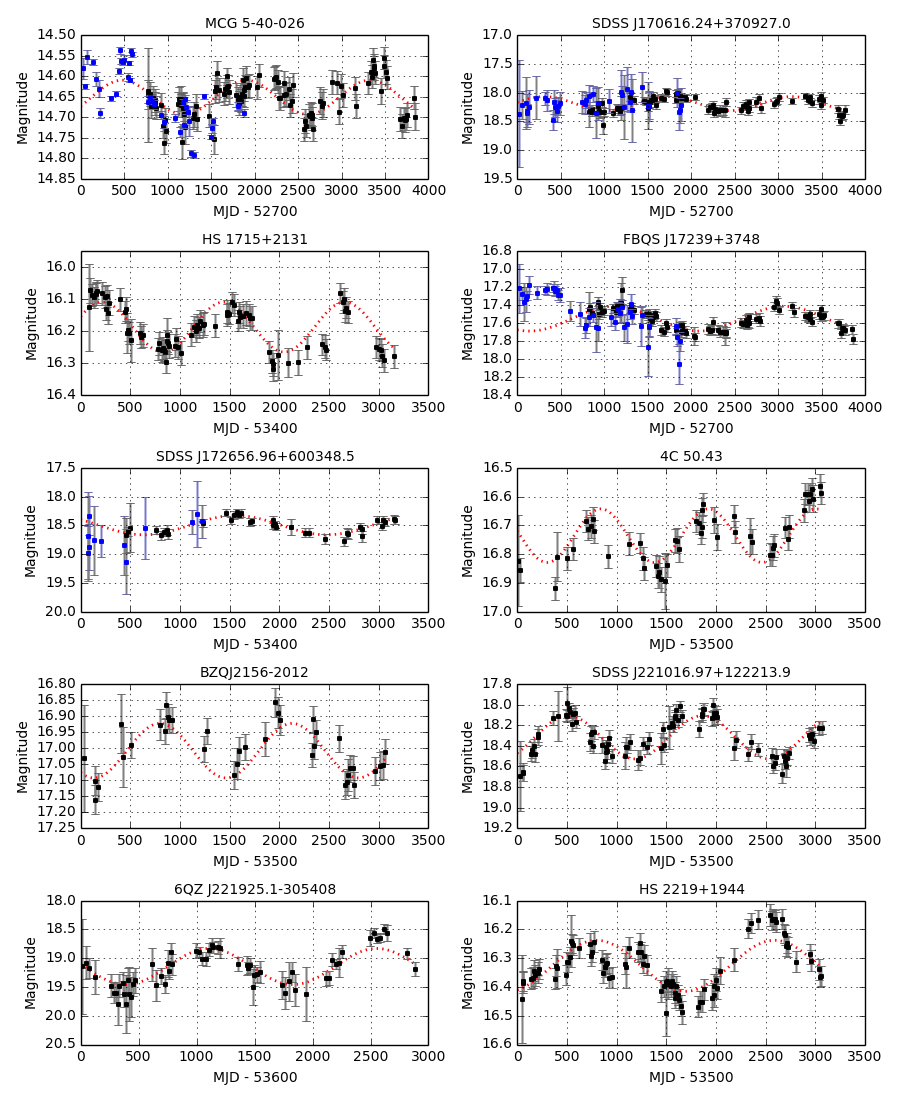} 
\end{figure*}

\begin{figure}
\centering
\contcaption{}
\includegraphics[width = 7.0in]{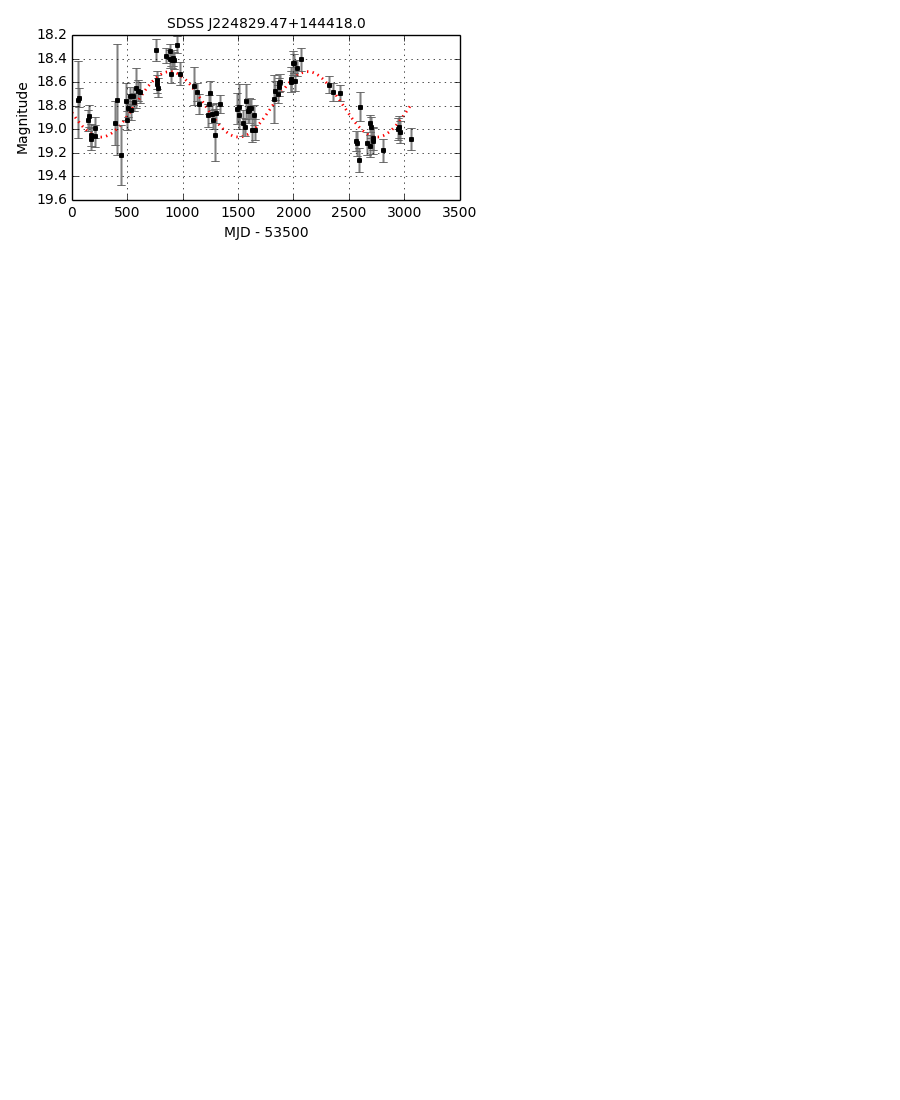} 
\end{figure}

\section{Binary Candidates in the Literature}

\subsection{Spectroscopic}

As previously mentioned, the availability of multiepoch spectroscopy enables searches for SMBH binaries on the basis of time-dependent velocity shifts in the broad lines, either with respect to each other or narrow lines associated with the host galaxy. It is interesting to see whether any spectroscopically selected candidates in the literature show periodic photometric variability in CRTS data. We have examined the light curves of 30 spectroscopic SMBH binary candidates from the literature (see Table~\ref{litcands}). We only consider here the most likely binary candidates reported, e.g., classified as such from the shape of their Balmer lines \citep{tsalmantza11, decarli13},  and not secondary candidates also reported with asymmetric line profiles, double-peaked emitters, or other features since these do not sufficiently indicate a SMBH binary. Although the predicted separation of these candidates is greater than in our candidates, it is still typically less than 1 pc and we would expect there to be some detectable behavior in the photometric data.

\begin{table*}
\caption{SMBH binary candidates identified in the literature}
\label{litcands}
\begin{tabular}{llllllllll}
\hline
 &  &  & \multicolumn{5}{c}{Best fits} & &  \\
Id & RA & Dec & Shape & Poly & $\log_{10} \tau$ & $\log_{10} \sigma^2$ & CARMA & MAD & Source \\
\hline
J0049+0026 & 00 49 19.0 & +00 26 09.4 & Hump & 3 & 2.999 & -4.579 & (2, 0) &  0.10 & Optical$^{10}$ \\
J0301+0004 & 03 01 00.2 & +00 04 29.3 & Periodic & 1 & -1.957 & 0.753 & (4, 0) & 0.21 & Optical$^{9}$ \\
J0322+0055 & 03 22 13.9 & +00 55 13.4 & Polynomial & 5 & 2.720 & -4.327& (6, 0) & 0.07 & Optical$^{9}$ \\
J0757+4248 & 07 57 00.7 & +42 48 14.5 & Logistic & 3 & 3.235 & -5.031 & (2, 0) &   0.11 & Optical$^{10}$ \\
J0829+2728 & 08 29 30.6 & +27 28 22.7 & Polynomial & 4 & 3.095 & -3.827 & (4, 0) & 0.23 & Optical$^{8,12}$ \\
J0847+3732 & 08 47 16.0 & +37 32 18.1 & Polynomial & 4 & 2.844 & -4.001 & (4, 0) & 0.12 & Optical$^{8}$ \\
J0852+2004 & 08 52 37.0 & +20 04 11.0 & Hump & 4 &  2.594 & -3.871 & (4, 0) &  0.15 & Optical$^{8}$ \\
OJ 287 & 08 54 48.9 & +20 06 30.6 & Periodic & 5 & 1.678 & -2.021 & (7, 0) & 0.34 & Optical$^{1}$  \\
J0927+2943 & 09 27 04.4 & +29 44 01.3 & Periodic & 2 & 2.882 & -4.639 & (1, 0) & 0.09 & Optical$^{2,7,11}$ \\
J0928+6025 & 09 28 38.0 & +60 25 21.0 & Periodic & 1 & 2.650 & -4.105 & (6, 0) & 0.10 & Optical$^{8}$ \\
J0932+0318 & 09 32 01.6 & +03 18 58.7 & Polynomial & 5 & 2.921 & -4.119 & (4, 0) & 0.13 & Optical$^{6,7,11}$ \\
J0935+4331 & 09 35 02.5 & +43 31 10.7 & Hump & 5 & 2.946 & -4.541 & (6, 0) & 0.05 & Optical$^{10}$ \\
4C 40.24	& 09 48 55.3 & +40 39 44.6  &  Polynomial & 5 & 2.900 & -4.500 & (2, 0) & 0.09 &Radio$^{13}$ \\
J0956+5350 & 09 56 56.4 & +53 50 23.2 & Periodic & 1 & 2.713 & -4.392 & (4, 0) & 0.10 & Optical$^{10}$ \\
J1000+2233 & 10 00 21.8 & +22 33 18.6 & Polynomial & 7 & 2.142 & -3.593 & (7, 0) & 0.12 & Optical$^{5,7,11}$ \\
J1012+2613 & 10 12 26.9 & +26 13 27.2 & Logistic & 2 & 2.936 & -4.557 & (5, 0) & 0.09 & Optical$^{7,11}$ \\
J1030+3102 & 10 30 59.1 & +31 02 55.8 & Hump & 3 & 2.780 & -4.134 & (5, 0) & 0.08 & Optical$^{8}$ \\
J1050+3456 & 10 50 41.4 & +34 56 31.3 & Polynomial & 10 & 2.689 & -4.154 & (2, 0) & 0.12 & Optical$^{4,7,11}$ \\
J1100+1709 & 11 00 51.0 & +17 09 34.3 & Hump & 1 & 2.906 & -4.232 & (6, 0) & 0.12 & Optical$^{8,12}$ \\
J1112+1813 & 11 12 30.9 & +18 31 11.4 & Polynomial & 6 & 2.784 & -4.264 & (3, 0) & 0.09   &  Optical$^{8}$ \\
J1154+0134 & 11 54 49.4& +01 34 43.6 & Logistic & 1 & 2.906 & -3.853 & (5, 2)  &  0.15 & Optical$^{7,11}$ \\
J1229-0035 & 12 29 09.5 & $-$00 35 30.0 & Polynomial & 4 & 2.834 & -3.947 & (6, 5) & 0.13 & Optical$^{9}$ \\
J1229+0203 & 12 29 06.7 & +02 03 08.7 & Polynomial & 5 & 1.397 & -3.057 & (7, 2) &  0.07 & Radio$^{13}$ \\
J1305+1819 & 13 05 34.5 & +18 19 32.9 & Polynomial & 8 & 2.728  & -4.899 & (5, 0) &  0.04 & Optical$^{8,12}$ \\
J1345+1144 & 13 45 48.5 & +11 44 43.5 & Hump & 4 & 2.440 & -4.039 & (6, 0) & 0.08 & Optical$^{8}$ \\
J1410+3643 & 14 10 20.6 & +36 43 22.7 & Logistic & 3 & 3.240 & -4.434 & (5, 0) &  0.15 & Optical$^{9}$ \\
J1536+0441 & 15 36 36.2 & +04 41 27.1 & Polynomial & 10 &  2.732 & -4.474 & (6, 0) & 0.07 & Optical$^{3,7,11}$ \\
J1537+0055 & 15 37 06.0 & +00 55 22.8 & Polynomial  & 2 & 3.105 & -5.987 & (2, 0) & 0.03 & Optical$^{9}$ \\
J1539+3333 & 15 39 08.1 & +33 33 27.6 &  Periodic & 1 & 3.042 & -5.591 & (2, 0) & 0.06 & Optical$^{7,11}$ \\
J1550+0521 & 15 50 53.2 & +05 21 12.1 & Hump & 1 & 2.804 & -4.817 & (2, 0) & 0.03 & Optical$^{9}$ \\
J1616+4341 & 16 16 09.5 & +43 41 46.8 & Logistic & 1 &-1.383 & -0.033 & (4, 0) & 0.18 & Optical$^{10}$ \\
J1714+3327 & 17 14 48.5 & +33 27 38.3 & Polynomial & 7 & 3.060 & -4.865 & (1, 0) & 0.06 & Optical$^{7,11}$ \\
PKS 2155-304 & 21 58 52.1 & $-$30 13 32.1 & Logistic & 1 & 2.483 & -2.648 & (3, 1) & 0.44 &  Radio$^{13}$\\
3C 454.3	& 22 53 57.7 & +16 08 53.6  & Polynomial & 4 & 1.978 & -1.817 & (6, 0) & 0.55 & Radio$^{13}$ \\
1ES 2321+419 & 23 23 52.1 & +42 10 59 & Polynomial & 5 & 2.531 & -2.974 & (6, 0) & 0.29 & X-ray$^{14}$ \\
J2349-0036 & 23 49 32.8 & $-$00 36 45.8 & Periodic & 1 &-0.337 & -1.153 & (7, 3) & 0.10 & Optical$^{9}$ \\
\hline
\end{tabular}

References: [1] Valtonen et al. (2008); [2] Dotti et al. (2009);  [3] Boroson \& Lauer 2009; [4] Shields, Bonning \& Salviander (2009)  ; [5] Decarli et al. (2010); [6] Barrows et al. (2011); [7] Tsalmantza et al. (2011); [8] Liu et al. (2014); [9] Shen et al. (2013); [10] \cite{ju13}; [11] \cite{decarli13}; [12] Eracleous et al. (2012); [13] Ciaramella et al. (2004); [14] Rani et al. (2009)
\end{table*}

There seem to be a number of morphologically distinct behaviors seen with these candidates. To quantify this, we have defined a number of template functions that capture different shape profiles:

\noindent
{\em Linear}: a line with gradient given by the Theil-Sen estimator for the light curve which passes through the median data point $(t_{med}, m_{med})$;

\noindent
{\em Polynomial}: the best fitting polynomial up to 10$^{th}$ order determined by a F-test between successive orders;

\noindent
{\em Sinusoidal}: the best fitting sinusoid with period determined by the ACF method (see section 2.2);

\noindent
{\em Logistic}: a generalized logistic function fit to the light curve:

\[ m(t) = A + \frac{K - A}{(1 + Q e^{-B(t-M)}) ^{1/\nu}} \]

\noindent
{\em Hump/Dip}: a linear least-squares fit to the light curve plus a quadratic to the largest contiguous amount of data above or below the linear fit.

For each light curve of interest, we have determined which template function provides the best chi-squared fit. The shape of the light curve may also just be a random pattern arising from the stochastic nature of the variability and so we have also determined both the best-fit CAR(1) and CARMA(p,q) model parameters for each light curve and checked for correlations between these and the best-fitting template. We note that none of the spectroscopic candidates shows the type of strong periodicity exhibited by the candidates identified in this paper. 

The division of template shapes is fairly uniform, although objects from \cite{liu14} tend to be best-fit by the hump template. These candidates specifically show significant radial accelerations in their broad H$\beta$ lines from dual epoch spectroscopy whereas other studies are based on more or other lines. There is also some suggestion of preferential localization in the CAR(1) $\tau-\sigma^2$ plane with objects best fit by a sinusoid function lying outside the $1\sigma$ contour. Those best fit by CARMA models with $p=7$  also show the same effect but to a higher degree. However, the sample size here is too small to determine whether these effects are statistically significant but they certainly warrant further study. 

\cite{fliu14} proposed SDSS J120136.02+300305.5 as the first milliparsec SMBH binary in a quiescent galaxy on the basis of a candidate tidal disruption event in its X-ray light curve. Its optical light curve shows no significant variation -- it has a MAD value of 0.07 mag which is the median value for its magnitude -- or evidence of an associated flare and there is no structure to it. This suggests that this population of SMBH binaries will only be detected in the optical as a result of serendipity (assuming that the candidate is an actual binary).

\subsection{Blazars}
Similarly we have examined the CRTS light curves of the few blazars reported in the literature as exhibiting \optprefix{quasi}{periodic} behavior,  e.g., PKS 2155-304 and 1ES 2321+419 \citep{rani09}. We note that these do not show the consistent periodicity that we see in the candidates in this analysis. However, their CAR(1) and CARMA fits are comparable with those seen for optical candidates in the literature as in the previous section.

\subsection{Photometric}
\cite{liu15} report the discovery of a periodic quasar candidate, FBQS J221648.7+012427, out of a sample of 168 quasars in the Pan-STARRS1 Medium Deep Survey (PSMDS) with an average of 350 detections in four filters. It has an observed period of 542 days, a redshift of 2.06, and an inferred separation of 0.006 pc. Whilst this object is part of the data we have considered, it does not appear as a candidate in our analysis. Assuming the quoted period, its CRTS light curve covers 5.8 cycles but no significant or consistent periodic signal is identified by either the WWZ or ACF algorithms and its Slepian wavelet characteristic timescale is not a statistically significant outlier.

The candidate was identified using a Lomb-Scargle (LS) periodogram to look for evidence of periodicity. A similar approach was used by \cite{mac10} in a search of 8863 SDSS Stripe 82 quasar light curves (with an average of over 60 observations) with 88 candidates identified. None were considered significant, however, as the standard likelihood calculation assumes white noise as the null hypothesis rather than colored noise as appropriate for an underlying CAR(1) process. The technique is regarded as not very powerful for poorly sampled light curves. It is also interesting to note that both analyses have a detection rate of about 1 in 100 rather than the far more conservative 1 in 10000 that we find, which is much more in line with expected rates from population models (see section~\ref{detection}).

The PSMDS candidate FBQS J221648.7+012427 is also reported to have a (restframe) inspiral time of $\sim$ 7 years thus putting its final coalescence within the timeframe of pulsar timing array experiments within the next decade or so. Given that the merger timescale is typically 10$^7$ years, it seems statistically unlikely that an object would be found in its final phase in such a small sample size. The inspiral time is, however, very dependent on the black hole mass and so it is possible that the authors have overestimated the mass of the system, particularly as the line and continuum fluxes were not determined from an electronic format spectrum.

\section{Discussion}

\subsection{Quasiperiodic behavior in stochastic models}
\label{carma}
Although CAR(1) models have been used as popular statistical descriptors of quasar optical variability, e.g., \cite{kelly09, mac12}, there is a growing body of evidence that a more sophisticated approach is required 
\citep{mushotzky11,zu13,graham14, kasliwal15}. \cite{kelly14} describe the use of CARMA models as a better method to quantify stochastic variability in a light curve. (Note that a CAR(1) model is the same as a CARMA(1, 0) model). A zero-mean CARMA(p,q) process $y(t)$ is defined to be the solution to the stochastic differential equation: 

\begin{eqnarray*}
\frac{d^py(t)}{dt^p} + \alpha_{p-1} \frac{d^{p-1}y(t)}{dt^{p-1}} + \ldots + \alpha_0y(t) = \\
\beta_q \frac{d^q\epsilon(t)}{dt^q} + \beta_{q-1} \frac{d^{q-1}\epsilon(t)}{dt^{q-1}} + \ldots + \epsilon(t)
\end{eqnarray*}

\noindent
where $\epsilon(t)$ is a continuous time white noise process with zero mean and variance $\sigma^2$. This is associated with a characteristic polynomial:

\[ A(z) = \sum_{k=0}^p \alpha_k z^k \]

\noindent
with roots $r_1, \ldots, r_p$. The power spectral density (PSD) for a CARMA(p, q) process is given by:

\[
P(\omega) = \sigma^{2} \frac{\mid \sum_{j=0}^q \beta_j(i \omega)^j \mid^2} {\mid \sum_{k=0}^p \alpha_k(i \omega)^k \mid^2}
\]

\noindent
Now $A(z)$ can be written in terms of its roots: 

\[ A(z) = \prod_{j=1}^p (z - r_j) = 0 \]

\noindent
where $r = a + ib$ and so the PSD is proportional to:

\[
P(\omega) \propto \frac{1}{\prod_{j = 1}^p a_j^2 + (w - b_j)^2}
\] 

\noindent 
For complex roots (and $p > 1$), local maxima will exist at $w = b_j$ which correspond to quasi-periodicities in the data. If the roots have negative real parts and $q < p$ then the CARMA process is stationary. By analogy with underdamped second order systems, the quasi-periodicities can be characterized by the damping ratio, $\zeta_j = -a_j / \sqrt{a_j^2 + b_j^2}$, for a given frequency, $b_j$. Smaller values of $\zeta$ indicate a stronger oscillating signal (less damping) with $\zeta = 0$ defining a pure sinusoidal signal. The bandwidth of the signal, i.e., the range of frequencies associated with it, is given by $\Delta f = \zeta b / \pi$. 

For each of our periodic candidates, we have determined the best-fitting CARMA(p,q) model using the {\tt carma\_pack} code\footnote{https://github.com/brandonckelly/carma\_pack} of \cite{kelly14} and checked whether there is any quasi-periodic behavior expected at its identified period from peaks in the PSD. Out of the 111 candidates, we find that five show a detected periodicity which overlaps with that expected from its best-fit CARMA
model. However, the bandwidth of all of these is such that any period within the temporal baseline of the object would fit. None of the periodic behavior is therefore consistent with that potentially expected from stochastic variability.

We have also generated 100 mock light curves of the appropriate CARMA(p,q) type for each periodic candidate with the same time sampling as the CRTS data and determined the expected wavelet and ACF parameters. The median differences between the real value and CARMA model-based values are statistically significant in all cases. From this we conclude that CARMA(p,q) models do not provide an adequate statistical description for  our periodic candidates as they cannot reproduce the same range of features values seen.

\subsection{Physical interpretation} 
Although we have searched for optical periodicity in quasars on the assumption that this is associated with a close SMBH binary system, the exact physical mechanism responsible for this behavior is uncertain. The sinusoidal nature of the signal suggests that it is kinematic in origin and there are a number of potential explanations that should be considered.

\subsubsection{A jet with a varying viewing angle}
The optical flux could be the superposition of thermal emission from the accretion disk and a differential Doppler boosted non-thermal contribution associated with a jet \citep{rieger04}. The observed periodicity would be the result of a periodically varying viewing angle, driven either by the orbital motion of a SMBH binary system, a precessing jet or an internally rotating jet flow. In the latter case, the expected observed period is typically $P_{obs} \lesssim 10$ days for massive quasars and even less for objects with significant bulk Lorentz factors, i.e., blazars. Given the range of values we have identified, this seems an unlikely model. 

In a single SMBH system, precession of the jet can arise if the accretion disk is tilted (or warped) with respect to the plane of symmetry (although the origin of the tilt would still need to be accounted for). Using data in the literature, \cite{lu05} find that the expected precession period for a jet with a single SMBH \citep{lu05} is:

\[ \log(\tau_{prec} / \mathrm{yr})  \sim 0.48M_{abs} + 0.14 \log \left( \frac{M}{10^8 M_\odot} \right) + A \]

\noindent
where $A = 15.18 - 18.86$, depending on the values of constant parameters relating to the viscosity and angular momentum of the SMBH system. For a representative quasar with $M_{abs} = -25$ with a $10^8 M_{\odot}$ SMBH, this gives a period between $10^{2.2 - 6.9}$ years which is much longer than the observed periods reported here. We note as well that jets in non-blazar systems are predominantly associated with geometrically thick accretion flows in low-luminosity AGN ($L_{bol} \le 10^{42}$ erg s$^{-1}$) 
whereas the candidate objects considered here span the range $10^{45} \le L_{bol} \le 10^{48}$ erg s$^{-1}$. Reported jetted systems in the high luminosity range tend to be blazars and radio galaxies, e.g., \cite{ghisellini10}, \cite{sbarrato14}, but only a few of the candidates fit that category. It is still possible, though, that the central part of the accretion disk in the highest luminosity systems could be geometrically thick with a slim disk associated to accretion rates close to or above the Eddington limit, e.g. \cite{sadowski15} and references therein.

Shorter periods are possible, however, for jet procession in a SMBH binary system. In this case, the jet precesses as a result of inner disk precession due to the tidal interaction of an inclined secondary SMBH.

\subsubsection{Warped accretion disk}
In addition to precessing any potential jet, a warped accretion disk might be responsible for the exhibited periodicity by obscuring the continuum-emitting region or at least modulating its luminosity as it precesses.
Warped features are known to exist when the data is of sufficient quality to identify them, e.g. \cite{herrnstein05} for NGC 4258, and may be common to most AGN, according to high resolution simulations of gas inflow in 
active galaxies \citep{hopkins10}. In general, the warping may be produced by a number of different mechanisms: the Bardeen-Peterson effect, a quadrupole potential or tidal interaction in a binary system, self-gravity, angular momentum transport by viscous disk stresses, and radiation- and magnetic-driven instabilities. 

Simulations suggest that the Bardeen-Peterson effect may not typically occur in AGN disks \citep{fragile07}. Radiation- and magnetic-driven instabilities are also not feasible for AGNs as they predict behaviors inconsistent with observation \citep{caproni06}. \cite{tremaine14} argue that, in the absence of any source of external torque, self-gravity is expected to play a prominent role in the dynamics of AGN accretion disks.
The precession rate of a self-gravitating warped disk around a SMBH is \citep{ulubay09}: 

\[ \omega \simeq C \left( \frac{G M_{d}^2}{M_{BH}  r_{w}^3} \right) ^ {\frac{1}{2}} \]

\noindent
where $r_w$ is the radius of the warp, $M_{BH}$ is the mass of the central object, $M_{d}$ is the mass of the accretion disk, and $C$ is a constant of order unity accounting for the disk configuration (inclination, etc). For a fiducial value of $M_{BH}$ = 10$^8 M_\odot$, \cite{tremaine14} find a warp radius of $\sim 300 R_g$ ($R_g = G M_{BH} / c^2$) and, assuming $M_{disk} = 0.01 M_{BH}$, the precession timescale is 51 years. This is only an order of magnitude larger than the typical periods reported here. The typical lifetime for a warp is $t_{align} = 1.3 \times 10^5 (r_w/ 300 R_g)^4$ yr \citep{tremaine14}, much shorter than the typical AGN lifetime, and so it would also be a relatively rare phenomenon to encounter.

A common explanation for a warped accretion disk is a binary system \citep{macfadyen08, hayasaki14, tremaine14}, particularly in stellar systems where all components are resolvable. In the case of a SMBH binary, a warp will occur in the accretion disk of one of the SMBHs if its spin axis is misaligned with the orbital axis of the binary. It is also possible that there is a circumbinary accretion disk which is warped.

\subsubsection{Periodic accretion}
\cite{farris14} describe how periodic mass accretion rates in a SMBH binary system can give rise to an overdense lump in the inner circumbinary accretion disk (CAD) and that the observation of periodicity in quasar emission would be associated with this. Periodic accretion would not be expected for a single SMBH system. \cite{hayasaki08} propose that small temperature changes associated with accretion variations would produce large-amplitude variations in the UV region, where the spectral energy distribution (SED) of the quasar decays exponentially, but would have little effect on the Rayleigh-Jeans part of the spectrum. However, \cite{rafikov13} argues that the SED of a circumbinary disk exhibits a power law segment with $\nu F_\nu \propto \nu^{12/7}$ rather than $\nu^{4/3}$. Accretion variations would therefore have a more noticeable effect at UV wavelengths in a binary system. We note as well that close pre-main-sequence binary stars show periodic changes in their observed optical luminosity which is attributed to periodic accretion from the circumbinary disk, e.g., \cite{jensen07}.

\subsubsection{Accretion disk gap}
From hydrodynamical simulations of SBMH binaries, \cite{dorazio15} have proposed that the detected periodicity in PG1302-102 could be associated with a lopsided central cavity in the CAD for $q > 0.3$, where $q$ is the ratio of the two SMBH masses: $q = M_2 / M_1$. Significantly, in this scenario, the strongest period detected is a factor of $\chi =  3 - 8$ times greater than the true period of the binary and so the true binary separation is a factor of $\chi^{-2/3}$ smaller. This model predicts additional smaller amplitude periodic variability on timescales of $t_{bin}$ and $0.5 t_{bin}$ ($t_{bin}$ is the true binary periodicity), and periodic spectral variations in broad line widths and narrow line offsets. There are also measurable relativistic effects on the Fe K$\alpha$ line and an observable decrement in the spectral energy distribution of the source at a wavelength determined by the width of the gap, e.g., \cite{gultekin12}.

\subsubsection{Relativistic beaming}
\cite{dorazio15} have also advanced an alternate suggestion for the periodicity of PG1302-102 -- the emission seen is from a mini-disk around the secondary black hole which is in orbital motion around the system barycenter. Doppler boosting of the emission as the disk moves along the line-of-sight is sufficient to account for the periodic variation seen. With the right choice of parameters, the variability of PG 1302-102 can explained by this if the broad lines do not originate from any circumbinary disk. The model also predicts different amplitude variability at different frequencies, assuming that the spectral slope is different, with no phase difference, i.e, it should track the optical variability.

\subsubsection{Quasi-periodic oscillation}
Quasi-periodic oscillations (QPOs) are a common phenomenon in the X-ray emission of stellar mass black hole binaries. Although the underlying physical cause is not understood, high frequency QPOs are consistent with the dynamical time of the system whilst low frequency ones are associated with Lense-Thirring precession of a geometrically thick accretion flow near the primary black hole, e.g., \cite{ingram09}. If the frequencies involved are scaled up by mass (many black hole timescales depend approximately on the inverse of the black hole mass), they are roughly consistent with the periodic behavior reported in this work. For example, the microquasar GRS 1915+105 has a mass of $\sim12 M_\odot$ and exhibits QPOs at $\sim$1 Hz (e.g., \cite{yan13} and references therein) with slowly varying frequency. Scaling this to the observed mass range of PG 1302-102 ($10^{8.3 - 9.4} M_\odot$) gives QPOs with a period between $\sim$200 and 2400 days (the restframe period is $\sim$1440 days).  \cite{king13} reported the first AGN analog of a low-frequency QPO in the 15 GHz light curve of the blazar J1359+4011 with a timescale varying between 120 and 150 days over a $\sim$4 year timespan. Stellar black hole binary QPOs are predicted to show a decreasing rms amplitude with longer wavelength \citep{veledina15}. The degree and angle of linear polarization of the precessing accretion flow are also predicted to modulate on the QPO frequency \citep{ingram15}.

\subsection{Theoretical detection rates}
\label{detection}
Given our large data set of  250,000 quasars, it is worthwhile considering how many SMBH binary systems we could expect to find. \cite{haiman09} used simple disk models for circumbinary gas and the binary-disk interaction to determine the number of SMBH binaries that may be expected in a variety of surveys, assuming that such objects are in the final gravitational wave-dominated phase of coalescence (which equates to separations less than $\sim$0.01 pc for a $10^8 M_\odot$ system). \cite{volonteri09} combined this with merger tree assembly models to similarly predict the number of expected SMBH binaries at wider separations where spectral line variations may be seen (equating to separations greater than $\sim 0.2$ pc for a $10^8 M_\odot$ SMBH binary). The latter shows that in a sample of 10000 quasars with $z < 0.7$, there should be $\sim10$ such objects and this number increases by a factor of $\sim$5--10 for $z < 1$. We note that our sample has $\sim 75000$ quasars with $z < 1$. 

Using these approaches, we can estimate our predicted sample size. Assuming a limiting magnitude of $V \sim 20$, a detectable range of orbital periods from 20--300 weeks (thus spanning both gravitational-wave and gas-dominated regimes), a survey sky coverage of $2\pi$ steradians (there are far fewer known spectroscopically confirmed quasars in the regions where CRTS does not overlap with SDSS), and a redshift range of 0.0 -- 4.5, 
we should identify $\sim 450$ SMBH binaries. Our detection rate of $10^{-4}$ is therefore conservative compared to the predicted rate of $2.3 \times 10^{-3}$.

Virial black hole mass estimates \citep{shen11} are available for 88,000 quasars in our sample (of which 23 per cent have $z > 2$). If we assume that each of these is a SMBH binary with a separation of 0.01 pc then the CRTS temporal baseline is sufficient to detect 1.5 cycles or more of periodic behavior in 63 per cent of them (including 55 per cent of the $z > 2$ population). Our search should therefore be sensitive to a large fraction of the close SMBH population. We note, however, these theoretical arguments are still subject to considerable uncertainties, for example, if the final parsec problem cannot be resolved then there will not be any binaries in the $\sim$0.01 pc regime.

\begin{figure}
\centering
\caption{The frequency of supermassive black hole binaries as a function of restframe period. This shows the distribution for objects with $\log(M_{BH} / M_{\odot}) < 9$ (blue) and $\log(M_{BH} / M_{\odot}) > 9$ (green). Theoretical curves are also shown for a $n_{var} \propto t_{orb}^{8/3}$ relationship for the two mass distributions: $\log(M_{BH} / M_\odot) < 9$ (black) and $\log(M_{BH} / M_\odot) > 9$ (red).}
\label{bhnum}
\includegraphics[width = 3.5in]{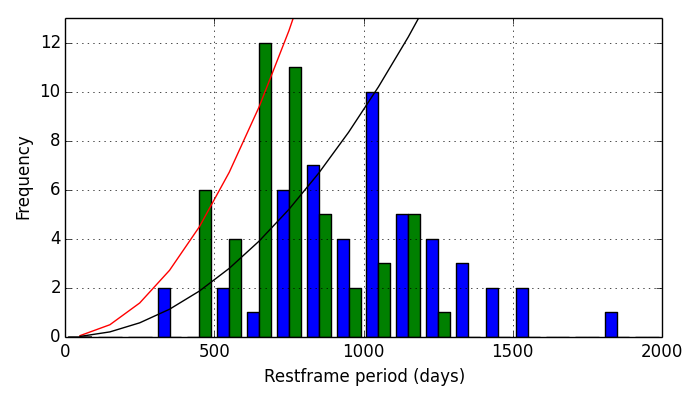}
\end{figure}

\cite{haiman09} also predict that in the gravitational wave (GW)-dominated regime of the merger process, the number of expected binaries is:

\[ n_{\mathrm var} = \left( \frac{10^7 \mathrm{yr}}{t_Q} \right) \left[ \frac{t_{\mathrm{orb}}}{50.2 \,\, \mathrm{week}} \right]^{8/3} M_7^{-5/3} q_s^{-1} 
\]

\noindent
where $t_{orb}$ is the restframe period, $t_Q$ is the timescale, $M_7$ is the black hole mass in units of $10^7 M_{\odot}$ and $q_s$ is the black hole mass ratio. Fig.~\ref{bhnum} shows the observed distributions for the binary candidates with $\log(M_{BH}/M_{\odot}) < 9$ and $\log(M_{BH}/M_{\odot}) > 9$ together with $n_{\mathrm var}  \propto t_{orb}^{8/3}$ 
relationships fit to these ($\log(M_{BH}/M_{\odot}) \sim 9$ is the median value for the data set and provides a natural division point). The results suggest that there is a transition timescale below which objects follow the power law expected from orbital decay driven by GWs. This timescale also anticorrelates with object mass such that 
higher mass objects (here $\log(M_{BH}/M_{\odot}) > 9$) change behavior at about $t_{orb} \sim 750$ days and lower mass objects at $t_{orb} \sim 1000$ days. 

At longer timescales (and lower masses), the number of objects is expected to follow a power law distribution $n_{var} \propto t^{\alpha}$, where the scaling index $\alpha$ is dependent on the physics of the circumbinary accretion disk and viscous orbital decay. However, since one of our selection criteria is that the temporal coverage of a light curve should cover at least 1.5 cycles of the periodic signal in the observed frame, we are incomplete at restframe timescales larger than the transition timescale. We can therefore not say anything about 
the true value of $\alpha$ and the physics of the gas-driven phase from our sample due to insufficient coverage. 
Nevertheless the statistical detection of GW-driven binaries can also be seen to confirm that circumbinary gas is present at small orbital radii and is being perturbed by the black holes \citep{haiman09}.

\section{Gravitational wave detection}
Given that some of the candidates appear to be in the gravitational-wave driven phase of merging, it is worth determining whether any of them might be resolvable as individual sources for current or upcoming nanohertz GW detectors (pulsar timing arrays (PTA)) rather than just be contributors to the stochastic GW background. For each candidate, we have determined the maximum signal that could be detected (see Table~\ref{candidates}). The GW frequency of a binary with a circular orbit is $f_{GW} = 2 / t_{per}$, where $t_{per}$ is the orbital period. The intrinsic GW strain amplitude is:

\[
h_s = \left( \frac{128}{15} \right)^{1/2} \frac{(GM_c)^{5/3}}{c^4 d_L} (\pi f)^{2/3}
\]

\noindent
where $M_c = (M_1M_2)^{3/5}(M_1 + M_2)^{-1/5}$ is the chirp mass and $d_L$ is the comoving distance to the source. The maximum induced timing residual amplitude is then $h_s / (2 \pi f_{GW})$. 

We have also calculated the inspiral time for each candidate (see Table~\ref{candidates}) to check that there is no significant frequency evolution over the lifetime of a detection experiment, i.e., $t_{expt} << t_{insp}$, where $t_{insp} = 5 (1+q)^2 a^4 / 2 q R_S^3$ and $R_S = 2GM/c^2$. The median inspiral time is $\sim$8,000 years, although there are four candidates predicted to merge within the next century and , with decadal baselines, we should be able to detect period changes in these objects.

\begin{figure}
\centering
\caption{The expected GW strain amplitude for the candidate sources. The error bars indicate the range of strain amplitudes for $q= 0.05$ to $q = 1$. The blue line indicates the all-sky averaged noise limit for the Nanograv experiment and the red line that of the most sensitive detection limit for the Parkes PTA.}
\label{gw}
\includegraphics[width = 3.5in]{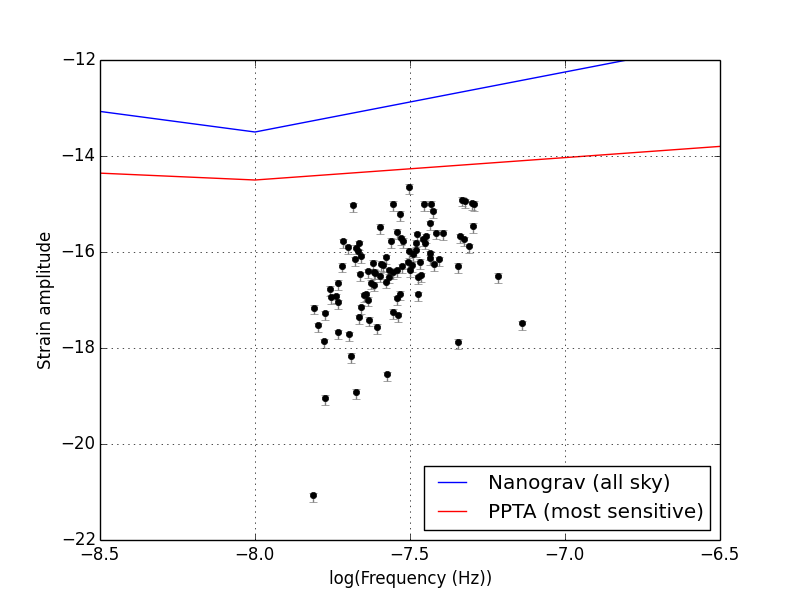}
\end{figure}

Using PTA noise estimates for Nanograv and the Parkes PTA (\cite{arzoumanian14,zhu14}), we calculate that none of the candidates will be resolvable as sources by the current generation of experiments (see Fig.~\ref{gw}). However, some of the sources will be viable by the next generation of detectors, e.g., SKA.

\section{Conclusions}

We have detected strong periodicity in the optical photometry of 111 quasars which we ascribe to the presence of a close SMBH binary in these systems. The Keplerian nature of the signal suggests that it is kinematic in origin and may be produced by a precessing jet, warped accretion disk or periodic accretion driven by the SMBH binary. We note our detection methodology may not be particularly sensitive to other types of periodic behavior exhibited by SMBH binary candidates, such as OJ 287. This would then suggest that our sample of objects represents a small fraction of a much larger close binary SMBH population which is yet to be detected. We are therefore planning further studies of the \optprefix{quasi}{periodic} quasar population with less stringent constraints than those applied in this analysis. This will also address our incomplete coverage of the gas-driven merger regime.

Further observations of our sample can test the different interpretations, as well as the primary SMBH binary attribution. \cite{shen10} suggest that reverberation mapping is a particularly useful diagnostic since the behaviour of emission line response to continuum variations is expected to be different for alternate explanations. We plan to see whether stacked reverberation employing time series and pairs of multiepoch spectra \citep{fine13} is sufficient. Obviously continued monitoring by CRTS and other synoptic surveys will track future cycles of periodicity and historical photometric data from photographic plate collections, such as DASCH\footnote{http://dasch.rc.fas.harvard.edu/project.php}, may provide more data for previous cycles. For example, DASCH data for PG 1302-102 is consistent with the reported period (J. Grindlay, private communication). With such decadal baselines, any change in the period of the system expected from general relativity may be detectable.

Future spectroscopic observations can also test whether there is any spectral variation in the sample consistent with binary orbital timescales, although this is not necessarily present in such close systems. The theoretical expectation for pairs at the close separations that we are probing is that they should not show any such effect -- the size of any broad line region (BLR) dwarfs the orbital dimensions of the binary and at separations closer than 0.03 pc in some scenarios, the BLR is truncated or destroyed \citep{ju13}. However, \cite{dorazio15} suggest that there may be broadened emission due to recombination in orbiting circumbinary gas which would show as periodic variation in the FWHM of associated lines and smaller shifts in their centroids. Continued spectral monitoring, particularly coincident with an extremum in the photometric light curve, will help to confirm the origin of any spectral variability detected. With a binary, the variation should follow a regular evolution whereas a bright spot (non-axisymmetric perturbation in the BLR emissivity) in the BLR, say, should just be a transient feature. One caveat with any detection of spectral variability is that asymmetric reverberation can act as a major source of confusion noise in multiepoch spectroscopic data \citep{barth15}.

Multiwavelength observations, particularly of the radio quiet objects, will also provide more information about them. In particular, the specific predictions that \cite{dorazio15} make regarding the Fe K$\alpha$ line, broad line widths and narrow line offsets for the accretion disc cavity model can be tested. Finally, we note that these objects are strong candidates for any gravitational wave experiment sensitive to nanohertz frequency waves such as those using pulsar timing arrays as well as any future space-borne gravitational wave detection mission.

\section*{Acknowledgments}

We thank Zoltan Haiman for useful discussions. 

This work was supported in part by the NSF grants AST-0909182, IIS-1118041 and AST-1313422, and by the W. M. Keck Institute for Space Studies. The work of DS was carried out at Jet Propulsion Laboratory, California Institute of Technology, under a contract with NASA.

This work made use of the Million Quasars Catalogue.

Funding for SDSS-III has been provided by the Alfred P. Sloan Foundation, the Participating Institutions, the National Science Foundation, and the U.S. Department of Energy Office of Science. The SDSS-III web site is http://www.sdss3.org/.

SDSS-III is managed by the Astrophysical Research Consortium for the Participating Institutions of the SDSS-III Collaboration including the University of Arizona, the Brazilian Participation Group, Brookhaven National Laboratory, Carnegie Mellon University, University of Florida, the French Participation Group, the German Participation Group, Harvard University, the Instituto de Astrofisica de Canarias, the Michigan State/Notre Dame/JINA Participation Group, Johns Hopkins University, Lawrence Berkeley National Laboratory, Max Planck Institute for Astrophysics, Max Planck Institute for Extraterrestrial Physics, New Mexico State University, New York University, Ohio State University, Pennsylvania State University, University of Portsmouth, Princeton University, the Spanish Participation Group, University of Tokyo, University of Utah, Vanderbilt University, University of Virginia, University of Washington, and Yale University.

\end{document}